\documentclass[12pt,12pt]{article}
\usepackage{graphicx}
\usepackage{epsfig}
\usepackage{amsmath,amssymb}
\usepackage{slashed} 
\font\mathbf cmbxti10 at 12pt

\catcode`\@=11
\newdimen\ex@
\font\dozeb=cmmib10 scaled \magstep1
\font\dozesyb=cmbsy10 scaled \magstep1
\font\dezb=cmmib10
\def\beq{\begin{equation}}
\def\eeq{\end{equation}}
\def\beqa{\begin{eqnarray}}
\def\eeqa{\end{eqnarray}}
\newcommand{\ba}{\begin{eqnarray}}
\newcommand{\ea}{\end{eqnarray}}
\newcommand\BA{\begin{array}}
\newcommand\EA{\end{array}}

\begin{document}
\def\thefootnote{\fnsymbol{footnote}}

\title{\bf Remarks on the Mean-Field Theory\\ 
Based on the SO(2N+1) Lie Algebra\\ 
of the Fermion Operators}
\vskip1cm
\author
{Seiya NISHIYAMA\footnotemark[1] $\!\!$, 
Jo\~ao da PROVID\^{E}NCIA\footnotemark[2]\\
\\
Centro de F\'\i sica,
Departamento de F\'\i sica,\\ 
Universidade de Coimbra\\
P-3004-516 Coimbra, Portugal\\
\\[-0.01cm]
{\it Dedicated to the Memory of Hideo Fukutome}}

\maketitle

\vspace{-0.5cm} 

\footnotetext[1]
{Corresponding author.
~E-mail address: seikoceu@khe.biglobe.ne.jp}
\footnotetext[2]
{E-mail address: providencia@teor.fis.uc.pt}

\begin{abstract}
Toward a unified algebraic theory 
for mean-field Hamiltonian (MFH) 
describing paired- and unpaired-mode effects,
in this paper, 
we propose a generalized HB (GHB) 
MFH
in terms of the $SO(2N \!\!+\!\! 1)$ Lie algebra of fermion pair and
creation-annihilation operators.
We diagonalize the GHB-MFH and 
throughout the diagonalization of which, 
we can first obtain the unpaired mode amplitudes which are given 
by the SCF parameters appeared in the HBT 
together with the additional SCF parameter in the GHB-MFH 
and by the parameter specifying the property of the SO(2N +1) group.
Consequently, it turns out that
the magnitudes of these amplitudes 
are governed by such parameters. 
Thus, it becomes possible to make clear a new aspect of such the results.
We construct the Killing potential
in the coset space $\frac{SO(2N)}{U(N)}$
on the K\"{a}hler symmetric space
which is equivalent with the generalized density matrix (GDM). 
We show another approach to
the fermion MFH based on such a GDM.
We derive an $SO(2N \!+\! 1)$ GHB MF operator and
a modified HB eigenvalue equation.
We discuss on the MF theory related to
the algebraic MF theory based on 
the GDM and the coadjoint orbit leading to
the non-degenerate symplectic form.
\end{abstract}
\vskip0.1cm

$Keywords$:

Mean-field Hamiltonian;

$SO(2N \!\!+\!\! 1)$ Lie algebra of fermion operators; 

Killing potential; 

Generalized density matrix;

Coadjoint orbit;



\vskip0.3cm


\vskip0.3cm



\newpage

\setcounter{equation}{0}
\renewcommand{\theequation}{\arabic{section}.\arabic{equation}}

\section{Introduction}


~~~
In nuclear and condensed matter physics,
Hartree-Bogoliubov theory (HBT)
\cite{BCS.57,Bog.59}
and time dependent HB theory (TDHBT)
\cite{RS.80,BR.86}
have been regarded as the standard approximation 
in many-body theoretical descriptions of 
superconducting fermion systems.
An HB wave function (WF) for such systems represents 
Bose condensate states of fermion pairs. 
It is a good approximation for the ground state of the system 
with a short-range pairing interaction
that produces a spontaneous Bose condensation of fermion pairs. 
Let us consider a fermion operator (OP) with $N$ single-particle states.
The pair OPs of the fermion form an $SO(2N)$ Lie algebra
and include a $U(N)$ Lie algebra as a sub-algebra. 
$\!$The $SO(2N)$ and the $U(N)$ denote 
the $2N$-dimensional special orthogonal group 
and the $2N$-dimensional unitary group, respectively.
One can give an integral representation of a state vector 
on the $SO(2N)$ group,
i.e., exact coherent state representation (CS rep) of a fermion system
\cite{Perelomov.86,Gilmore.74}.
The canonical transformation (TR) of the fermion OPs generated by 
the Lie OPs of the $SO(2N)$ Lie algebra induces 
the well-known generalized fermion Bogoliubov TR.
One of the present authors (S. N.) has shown that
a quantized TDHBT based on the $SO(2N)$ group
is obtained  by using the path integral
on the coset space
$\frac{SO(2N)}{U(N)}$.
The lowest order approximation in the path integral
leads to
a classical Euler-Lagrange equation 
of motion for the
$\frac{SO(2N)}{U(N)}$ coset variables.
From this, we can get the TDHB equation 
\cite{Ni.81}.
Usually, it solutions gives the ground state of
an even fermion system.
For an odd system, 
we have no theoretical method describable both
paired and unpaired states in an equal manner.

While,
in elementary particle physics,
supersymmetric extension of nonlinear models 
was first given by Zumino introducing scalar fields
taking value in a complex K\"{a}hler manifold
\cite{Zumino.79}.
The nonlinear $\sigma$-models 
defined on symmetric spaces
have been intensively studyied.
van Holten
gave a consistent coupling of gauge- and 
matter superfield to supersymmetric $\sigma$-model 
on the K\"{a}hler coset spaces
and provided 
the Killing potential and then
the explicit construction of the $\sigma$-model
on the coset space $\frac{SO(2N)}{U(N)}$. 
Such a coset $\sigma$-model is shown to consistently incorporates 
a matter in the representation
descending from the spinorial representations 
of the $SO(2N)$
\cite{NNH.01}.

It is still a current problem to give a theory
suitable for description of collective
motions with large amplitudes in nuclei with
strongly collective correlations. 
For a consistent description 
of collective excitations in such fermion systems,  
Fukutome, Yamamura and S. N. have proposed
a new fermion many-body theory
based on the $\!SO(2N \!+\! 1)\!$ Lie algebra of the fermion OPs 
\cite{FYN.77}.
The set of the fermion OPs 
composed of creation-annihilation and pair OPs forms a larger Lie algebra, 
Lie algebra of the $SO(2N \!+\! 1)$ group.
A representation of the $SO(2N \!+\! 1)$ group has been derived 
by a group extension of the $SO(2N)$ Bogoliubov TR 
for the fermions to a new canonical TR group
\cite{Sch.65,Fuk.77,Dob.82}. 
The fermion Lie OPs, 
when operating on the integral representation of the $SO(2N\!+\!1)$ WF, 
are mapped into the regular representation of 
the $SO(2N\!+\!1)$ group and are represented by some bosonized OPs. 
The boson images of the fermion Lie OPs are expressed by 
the closed first order differential forms.
In the regular representation space,
a classical equation of motion, i.e.,
the $SO(2N\!+\!1)$ TDHB equation which
describes the motion of fermionic $(2N\!+\!1)$-dimensional rotator,
has also been presented.Basing on such the theory of the fermionic rotator,
Fukutome and S. N.
have proposed the $SO(2N\!+\!1)$ TDHBT
for unified description of bose and fermi type collective excitations
\cite{FukNishi.84}.
S. N.
has also proposed another type of the $SO(2N\!+\!1)$ TDHB equation$\!$
\cite{Ni.82}.
As the $SO(2N \!+ 2)$ Lie OPs,
operated onto functions on the
$\!\frac{SO(2N+2)}{U(N+1)}\!$
coset manifold,
are mapped into the regular representation consisting of 
those functions,
we have reached an extended TDHBT (ETDHBT)
on the coset space
$\!\frac{SO(2N+2)}{U(N+1)}\!$
\cite{Nishi.98}.
Embedding the $SO(2N\!+\!1)$ group into 
an $SO(2N\!+\!2)$ group
and using the boson images of the fermion Lie OPs, 
we have obtained a new ETDHBT for
fermionic $(2N\!+\!2)$-dimensional rotator. 

$\!\!$Following a fundamental prescription prepared 
in the fermion $SO(2N \!+\!1)$ many-body theory,
the $SO(2N)~\sigma$-model has been extended 
by an algebraic manner to 
an $SO(2N \!+\! 1)~\sigma$-model
by the present authors $et~al.$
\cite{NishiProviCord.08,NishiProviCord.11}. 
Through the minimization of a scalar potential,
we have first determined a symbolical group-parameter $z^2\!$
specifying the property of the $SO(2N\!+\!1)$ group. 

$\!\!$The above new ETDHBT, of course, starts 
from the  original two-body fermion Hamiltonian. 
However, it involves the unknown parameters which originate 
from the Lagrange multipliers 
in order to select out the physical spinor subspace. 
Because it describes the fermionic $(2N\!\!+\!\!1)$-dimensional rotator,
through the coordinate transformations for the space fixed and 
the body fixed coordinate frames, 
the fermion $SO(2N+1)$ Lie OPs are expressed
in terms of the quasi-particle expectation values ($c$-number)
 of those Lie OPs 
and the quasi-particle $SO(2N\!\!+\!\!1)$ Lie OPs 
(quantum mechanical fluctuations).
The unknown parameters cannot be determined in the classical limit only, 
and a complete determination of them requires that 
the quantum mechanical fluctuations are taken into account. 
Instead, we have tried to determine the parameters with the aid of 
the quasi anti-commutation relation approximation 
for the fermion OPs. 
A determination of the parameters is possible 
if we demand that expectation values of 
the anti-commutators by an $SO(2N\!+\!1)$ HB WF 
satisfy the anti-commutation relations in the classical limit,
i.e., the quasi anti-commutation relation approximation for the fermion OPs 
\cite{Nishi.98}.
Under
the approximation, the determination has been attempted but 
has unfortunately been executed incompletely
\cite{NishiProvi.12}.
To approach such the problem of a fermion system,
theory for mean field approximation (MFA)
is the standard approximation  prevailed in the wide range of physics
and it is made by a manner of self-consistent field (SCF).
The new static EHBT
is derived from the new ETDHBT
\cite{NishiProvi.12}.
It is also a SCF theory
and applicable to both even and odd fermion systems 
which have the same capacity to provide
a MFA as the usual HBT for even fermion one.
The new EHB equation is written in terms of 
variables of paired and unpaired modes. 
Assuming a pairing potential,
it has been solved by a method parallel to the
two-step diagonalization method for the usual HB eigenvalue equation
\cite{Baranger.60,Belyaev.68}. 
We have obtained a new solution with unpaired-mode effects
arising from the Lagrange multipliers. 
We, however, have no effective SCF theory for the determination of
the unknown parameters in those multipliers. 
Thus for a long time it has not
been known yet how to describe self-consistent processes 
involving unpaired fermions.
In this sense,
we have no TD SCF theory for the description of 
both the paired and unpaired modes. 

Toward a unified algebraic theory 
for the MFA describing the paired- and unpaired-mode effects,
in this paper, 
we will propose a generalized HB (GHB) 
mean-field Hamiltonian (MFH)
in terms of the $SO(2N \!\!+\!\! 1)$ Lie algebra of fermion pair and
creation-annihilation OPs.
This GHB-MFH is very similar to the one
linear in the Jacobi generator for boson system
\cite{NishiProvi.18}.
We diagonalize the GHB-MFH and 
throughout the diagonalization of which, 
we can first obtain the unpaired mode amplitudes 
which are given by the SCF parameters appeared in the HBT
together with the additional SCF$\!$ parameter$\!$
in the GHB-MFH$\!$
and$\!$ by$\!$ the $\!$group-parameter $\!z$.
$\!\!\!$Consequently,
it turns out that
the magnitudes of the amplitudes are governed by
such parameters.
Thus,
it becomes possible to make clear a new aspect of such the results.
We construct a Killing potential
in the coset space $\!\frac{SO(2N)}{U(N)}\!$.
To our great surprise,
the Killing potential is equivalent with the generalized density matrix (GDM). 
We show another approach to
the fermion MFH based on
such a GDM.
We derive an $\!SO(2N \!\!+\!\! 1)\!$ GHB MF OP and
a modified HB eigenvalue equation (EE).
At last
we can give a new fermion MFT
on the K\"{a}hler coset space
$\frac{SO(2N + 2)}{U(N + 1)}$.

$\!$In \S 2,
we give a brief review on 
the $SO(2N \!+\! 1)$ Bogoliubov TR
and
the Killing potential equivalent with the GDM.
In \S 3,
on the $SO(2N \!+\! 1)$ group manifold, 
we give a GHB MFH
and diagonalization of the GHB MFH.
A  new aspect of the diagonalized solution
is made clear.
In \S 4,
MF approach using another form of the GDM
is robustly developed.
Finally, in \S 5, 
we give some 
discussions on the coadjoint orbit and the symplectic form,
perspective and summary.


\newpage

\setcounter{equation}{0}
\renewcommand{\theequation}{\arabic{section}.\arabic{equation}}

\section{SO(2N+1) Bogoliubov transformation
and GDM}

\def\bra#1{{< \!\!#1|}} 
\def\ket#1{{| #1 \!\!>}}

~~~~Let $c_{\alpha }$ and $c^{\dag }_{\alpha }$, $\alpha$ 
\!=\! 
1,$\cdot \cdot \cdot$, 
$N$, be annihilation and creation OPs of the fermion system
satisfying 
the canonical anti-commutation relations
$
\{c_{\alpha },c^{\dag }_{\beta }\}
\!=\!
\delta_{\alpha \beta } ,
\{c_{\alpha },c_{\beta }\}
\!=\!
\{c^{\dag }_{\alpha },c^{\dag }_{\beta }\}
\!=\!
0
$ .
We introduce a set of fermion OPs consisting of the
annihilation and creation OPs and pair OPs,\\[-18pt] 
\beqa
\BA{l}
c_{\alpha },~c^{\dag }_{\alpha } ,~~
E^{\alpha }_{~\beta }
\!=\!
c^{\dag }_{\alpha }c_{\beta }
{\displaystyle -\frac{1}{2}} \delta_{\alpha \beta }
\!=\!
E^{\beta \dag }_{~\alpha } ,~~
E^{\alpha \beta }
\!=\!
c^{\dag }_{\alpha }c^{\dag }_{\beta }
\!=\!
E^{\dag }_{\beta \alpha } ,~~
E_{\alpha \beta }
\!=\!
c_{\alpha }c_{\beta }
\!=\!
- E_{\beta \alpha } ,
\EA
\label{operatorset}
\eeqa\\[-12pt]
which form the $SO(2N+1)$ Lie algebra.
We here omit the explicit expressions for
the commutation relations between 
their OPs.
The $SO(2N+1)$ Lie algebra of the fermion OPs contains
the $U(N)(\!=\!\{E^{\alpha }_{~\beta }\}\!)$ 
and 
the $SO(2N)(\!=\!\{E^{\alpha }_{~\beta },
E^{\alpha \beta },E_{\alpha \beta }\}\!)$ 
Lie algebras of the pair OPs 
as sub-algebras.

A unitary operator $U(g)$ deriving an $SO(2N)$ canonical TR 
is generated by 
the $SO(2N)$ Lie OPs.
The $U(g)$ induces the generalized Bogoliubov TR 
\cite{Bog.59} 
specified by an $SO(2N)$ matrix $g$,\\[-6pt]
\beq
U(g)(c, c^{\dag })U^{\dag }(g)
\!=\!
(c, c^{\dag }) g ,
\label{Bogotrans}
\eeq
\beq
g
\stackrel{\mathrm{def}}{=}
\left[ \!\!
\BA{cc} 
a & \overline{b} \\
b & \overline{a} \\ 
\EA \!\!
\right] ,
~
g^{\dag }g \!=\! gg^{\dag } \!=\! 1_{2N} ,~
\det g
\!=\!
1 ,~
(\det:\mbox{determinant}),
\label{RepBogotrans}
\eeq
\vspace{-0.2cm}
\beq
U(g)U(g') \!=\! U(gg') ,~~~
U(g^{-1}) \!=\! U^{-1}(g) \!=\! U^{\dag }(g) ,~~~
U(1_{2N}) \!=\! \mathbb{I}_g~(\mbox{unit operator on}~g),
\label{Ug}
\eeq\\[-10pt]
where ($c$, $c^{\dag }$) is 
the 2$N$-dimensional row vector 
(($c_{\alpha }$), ($c^{\dag }_{\alpha }$)) and 
$a \!=\! (a^{\alpha }_{~\beta })$ and $b \!=\! (b_{\alpha \beta })$ 
are $N \!\!\times\!\! N$ matrices. 
The bar denotes the complex conjugation
and the symbol {\scriptsize T} the transposition. 
The HB WF $\ket g$, $SO(2N)$ CS rep, is generated as 
$\ket g \!=\! U(g) \ket 0$,  
$\ket 0$:vacuum satisfying 
$c_{\alpha }\ket 0 \!=\! 0$. 

Let $n$ be the number OP 
$n \!=\! c^\dag _{\alpha } c_\alpha$.
The operator $(-1)^n$ anticommutes with 
$c_\alpha$ and $c^\dag _\alpha$;\\[-10pt]
\beq
\{ c_\alpha,~(-1)^n \}
\!=\!
\{ c^\dag _\alpha,~(-1)^n \}
\!=\!
0.
\label{chiralop}
\eeq\\[-10pt]
Introduce the OP
$
\Theta
$\!
(
$ \!
\Theta
\!\equiv\!
\theta_\alpha c^\dag_\alpha \!-\! \overline{\theta }_\alpha c_\alpha 
$).
Due to the relation
$
\Theta ^2
\!=\!
-
\overline{\theta }_\alpha \theta_\alpha
\!\equiv\!
-
\theta ^2
$,
we have\\[-18pt]
\beqa
\BA{l}
e^\Theta
\!=\!
Z 
\!+\! 
X_\alpha c^\dag_\alpha
\!-\! 
\overline{X}_\alpha c_\alpha ,~
\overline{X}_\alpha X_\alpha \!+\! Z^2 
\!=\! 
1 ,~
Z
\!=\!
\cos \theta ,~
X_\alpha
\!=\!
{\displaystyle \frac{\theta_\alpha }{\theta }} \sin \theta .
\EA
\label{theta}
\eeqa\\[-16pt]
From
(\ref{chiralop}) and (\ref{theta}),
we obtain\\[-18pt]
\beqa
\!\!\!\!
\BA{ll}
e^\Theta (c, c^\dag ,
{\displaystyle \frac{1}{\sqrt{2}}}) (\!-1\!)^n e^{-\Theta }
\!=\!
(c, c^\dag ,
{\displaystyle \frac{1}{\sqrt{2}}}) (\!-1\!)^n G_X,~ 
G_X 
\!\stackrel{\mathrm{def}}{=}\!
\left[ \!\!\!
\BA{ccc} 
I_N 
\!\!-\!\!
\overline{X} X^{\mbox{\scriptsize T}} &\!\!\!\!
\overline{X} X^\dag &\!\!\!\! -\sqrt{2}Z \overline{X} \\
\\[-8pt]
X X^{\mbox{\scriptsize T}} &\!\!\!\! I_N 
\!\!-\!\! 
X X^\dag &\!\!\!\! 
\sqrt{2}Z X  \\
\\[-8pt]
\sqrt{2} Z X^{\mbox{\scriptsize T}} &\!\!\!\! -\sqrt{2} Z X^\dag &\!\!\!\! 
2Z^2 \!\!-\!\! 1 
\EA \!\!\!
\right] \! .
\EA
\label{chiraloptrans}
\eeqa\\[-8pt]
Let $G$ be the $(2N+1) \times (2N+1)$ matrix defined by\\[-10pt]
\beqa
\!\!\!\!\!\!\!\!
\left.
\BA{ll}
&G
\!\equiv\!
G_X \!
\left[ \!
\BA{ccc} 
a & \overline{b} & 0 \\
\\[-8pt]
b & \overline{a} & 0 \\
\\[-8pt]
0 & 0 & 1
\EA \!
\right]
\!=\!  
\left[ \!
\BA{ccc} 
a - \overline{X} Y & \overline{b} + \overline{X} \overline{Y} &
-\sqrt{2}Z \overline{X} \\
\\[-8pt]
b + X Y & \overline{a} - X \overline{Y} & \sqrt{2}ZX  \\
\\[-8pt]
\sqrt{2}ZY & -\sqrt{2}Z \overline{Y} & 2Z^2 - 1
\EA \!
\right] ,
\BA{c}
X
\!=\!
\overline{a} Y^{\mbox{\scriptsize T}} 
\!-\! 
b Y^\dag ,\\
\\[-8pt]
Y
\!=\!
X^{\mbox{\scriptsize T}} a 
\!-\! 
X^\dag b ,\\
\\[-8pt]
Y Y^\dag \!+\! Z^2 = 1 .
\EA
\EA \!\!
\right\}
\label{defG} 
\eeqa\\[-6pt]
The $X$ and $Y$ are the column vector and the row vector,
respectively.
The $SO(2N\!+\!1)$ canonical TR $U(G)$ is generated by 
the fermion $SO(2N\!+\!1)$ Lie OPs.
The $U(G)$ is an extension of 
the generalized Bogoliubov TR $U(g)$ 
\cite{Bog.59} 
to a nonlinear Bogoliubov TR.

From
(\ref{Bogotrans}), (\ref{chiraloptrans}) and (\ref{defG})
and the commutability of $U(g)$ with $(-1)^n$,
we obtain\\[-14pt]
\beqa
\!\!\!\!
\BA{c}
U(G)(c, c^{\dag } ,{\displaystyle \frac{1}{\sqrt{2}}}) (-1)^n U^{\dag }(G)
\!=\!
(c, c^{\dag }, {\displaystyle \frac{1}{\sqrt{2}}} (-1)^n G,~~
G 
\!\stackrel{\mathrm{def}}{=}\!
\left[ \!\!
\BA{ccc} 
A &\!\!\!\! \overline{B} &\!\!\!\! {\displaystyle -\frac{\overline{x}}{\sqrt{2}}} \\ 
\\[-14pt]
B &\!\!\!\! \overline{A} &\!\!\!\! {\displaystyle \frac{x}{\sqrt{2}}} \\
\\[-14pt]
{\displaystyle \frac{y}{\sqrt{2}}} &\!\!\!\! 
{\displaystyle -\frac{\overline{y}}{\sqrt{2}}} &\!\!\!\! z 
\EA \!\!
\right] \! ,
\label{SO2Nplus1chiraltrans}
\EA  
\eeqa
where
$N \!\times\! N$ matrices
$A \!=\! (A^{\alpha }_{~\beta })$ 
and 
$B \!=\! (B_{\alpha \beta })$  
and
$N$-dimensional
column and row vectors
$x \!=\! (x_{\alpha })$ 
and 
$y \!=\! (y_i)$ 
and
$z$
are defined
as follows:\\[-16pt]
\beqa
\!\!\!\!
\BA{l}
A
\!\equiv\!
a
\!-\!
\overline{X} Y
\!=\!
a
\!-\!
{\displaystyle \frac{\overline{x} y}{2(1 \!+\! z)}} ,~
B
\!\equiv\!
b
\!+\!
X Y
\!=\!
b
\!+\!
{\displaystyle \frac{x y }{2(1 \!+\! z)}} ,~
x
\!\equiv\!
2Z X ,~
y
\!\equiv\!
2Z Y ,~
z
\!\equiv\!
2Z^2 \!-\! 1 .
\EA 
\label{relAtoaXY}
\eeqa\\[-12pt]
By using the relation 
$
U(G)(c, c^{\dag },{\displaystyle \frac{1}{\sqrt{2}}}) U^{\dag }(G)
\!=\!
U(G)(c, c^{\dag },{\displaystyle \frac{1}{\sqrt{2}}}) U^{\dag }(G)
(z \!+\! \rho)(-1)^n
$ 
and
the third column equation of
(\ref{SO2Nplus1chiraltrans}),
Eq.
(\ref{SO2Nplus1chiraltrans})
can be written 
with a $q$-number gauge factor $(z \!-\! \rho)$
\cite{FYN.77}
as\\[-6pt]
\beq
U(G)(c, c^{\dag },{\displaystyle \frac{1}{\sqrt{2}}}) U^{\dag }(G)
\!=\!
(c, c^{\dag }, {\displaystyle \frac{1}{\sqrt{2}}}) 
(z \!-\! \rho)G ,~~
G^{\dag }G \!=\! GG^{\dag } \!=\! 1_{2N+1} ,~~
\det G
\!=\!
1 ,
\label{UGdetG}
\eeq
\vspace{-0.1cm}
\beq
U(G)U(G') \!=\! U(GG') ,~~~
U(G^{-1}) \!=\! U^{-1}(G) \!=\! U^{\dag }(G) ,~~~
U(1_{2N+1}) \!=\! \mathbb{I}_G ,
\label{UGUG'}
\eeq\\[-10pt]
The $U\!(G)$ is the nonlinear TR 
with the $q$-number gauge factor $(z \!-\! \rho)$ 
where
${\rho}$
is given as
${\rho} 
\!=\! 
x_{\alpha }c^{\dag }_{\alpha }
\!-\!
\overline{x}_{\alpha }c_{\alpha }
$ 
and 
$
{\rho }^{2} 
\!=\! 
- 
\overline{x}_{\alpha }x_{\alpha } 
\!=\! 
{z}^{2} 
\!-\! 
1
$ 
\cite{Fuk.81,Dob.81I,Dob.81II}.

Let us introduce a $2N \!\times\! N$ isometric matrix 
$u$ by
$
u^{\mbox{\scriptsize T}}
\!\!=\!\!
\left[  
b^{\mbox{\scriptsize T}},  a^{\mbox{\scriptsize T}} 
\right] 
$
and an $\frac{SO(2N)}{U(N)}$ coset variable $q$ as
$q \!=\! ba^{-1} \!=\! -q^{\mbox{\scriptsize T}}$.
According to Zumino
\cite{Zumino.79},
if $a$ is non-singular,
we have relations governing $u^\dag u$ as\\[-16pt]
\beqa
\!\!\!\!
\BA{l}
u^\dag u
\!\!=\!\!
a^\dag a
\!\!+\!\!
b^\dag b
\!=\!
a^\dag \!\!
\left( \!
1_{\!N} 
\!\!+\!\! 
q^\dag q
\right) \!
a ,~
\ln \det u^\dag u
\!\!=\!\!
{\cal K}(q^\dag\!, q)
\!\!+\!\!
\ln \det a
\!\!+\!\!
\ln \det a^\dag \! ,
({\cal K}\!:\!\mbox{K\"{a}hler potential}) .
\EA
\label{lndetUUdagger}
\eeqa\\[-16pt]
Following van Holten
\cite{NNH.01},
we introduce $N$-dimensional matrices 
${\cal R}(q; \delta g)$, 
${\cal R}_T(q; \delta g)$ and $\chi$ by\\[-16pt]
\beqa
\BA{l}
{\cal R}(q; \delta g)
\!=\!
\delta b
\!-\! q \delta \overline{b} q
\!-\! q \delta a 
\!+\! \delta \overline{a} q  ,~~
{\cal R}_T (q; \delta g)
\!=\!
\delta \overline{a}
\!-\! 
q \delta \overline{b} ,~~
\chi
\!=\!
(1_{\!N} \!+\! q q^\dag)^{-1}
\!=\! 
\chi^\dag .
\EA
\label{RRTChi}
\eeqa\\[-16pt]
Then the Killing potential ${\cal M}$ is given as\\[-16pt]
\beqa
\BA{l}
i{\cal M} \!
\left(
q, \overline{q};\delta g
\right)
\!\stackrel{\mathrm{def}}{=}\!
\mbox{tr} \!
\left\{ \!
{\cal R}_T (q; \delta g)
\!-\!
{\cal R}(q; \delta g) q^\dag \chi \!
\right\}
\!=\!
\mbox{tr} \!
\left\{ \!
\left(
q \delta a q^\dag
\!+\! 
\delta \overline{a}
\!-\!
\delta b q^\dag 
\!-\! 
q \delta \overline{b}
\right) \!
\chi \!
\right\} .
\EA 
\label{formKillingpotM} 
\eeqa\\[-14pt]
Using
(\ref{formKillingpotM}),
each component of the expression for ${\cal M}$ 
is obtained as\\[-10pt]
\beq
i{\cal M}_{ \delta a}
\!=\!
q^\dag \chi q ,~~~~
i{\cal M}_{ \delta \overline{a}}
\!=\!
\chi ,~~~~
i{\cal M}_{\delta b}
\!=\!
-\chi q ,~~~~
i{\cal M}_{ \delta \overline{b} }
\!=\!
-q^\dag \chi .
\label{componentKillingpotM} 
\eeq\\[-12pt]
To make clear the meaning of the Killing potential,
using the $2N \!\times\! N$ isometric matrix 
$u (u^\dag \! u \!\!=\!\! 1_{N})$,
let us introduce the following 
$2N \!\times\! 2N$ matrix:\\[-2pt]
\beq
W
\!=\!
\overline{u}
u^{\mbox{\scriptsize T}}
\!=\!
\left[ \!
\BA{cc} 
R & -\overline{K} \\
\\[-4pt]
K & 1_{N} - \overline{R} \!
\EA
\right] \! ,
\BA{l}
R
\!=\!
\overline{b} b^{\mbox{\scriptsize T}}
\!=\!
q^\dag (1_{\!N} \!\!+\!\! q q^\dag )^{-1} q
\!=\!
q^\dag \chi q
\!=\!
1_{\!N} \!\!-\!\! \overline{\chi } , \\
\\[-4pt]
K
\!=\!
\overline{a}b^{\mbox{\scriptsize T}}
\!=\!
-
(1_{\!N} \!\!+\!\! q q^\dag )^{-1}q
\!=\!
-
\chi q  ,
\EA
~~
W^2 \!=\! W .
\label{densitymat}
\eeq\\[-2pt]
The $W$ is  
the generalized density matrix (GDM) in the $SO(2N)$ CS rep,
with $R$ and $K$ in
(\ref{expectG}).\\[-14pt]

To  exploit a unified description of
the unpaired-mode solved amplitudes
obtained in the previous section and
the paired mode amplitudes,
we use another form of GDM
\cite{FukNishi.84} 
defined as\\[-2pt]
\beq
\slashed{W}
\!=\!
\left[ \!
\BA{cc} 
2 R - 1_{\!N} &\!\! - 2 \overline{K} \\
\\[-2pt]
2 K &\!\! - 2 \overline{R} + 1_{N}
\EA \!
\right]
\!=\!
g \!
\left[ \!
\BA{cc} 
-1_{N} &\! 0 \\
\\[-2pt]
0 &\! 1_{N}
\EA \!
\right] \!
g^\dag ,
~~
\slashed{W}^\dag
\!=\!
\slashed{W},
~~
\slashed{W}^2 \!=\! 1_{2N} .
\label{anotherGDM}
\eeq\\[-1pt]
Referring to the form of the above GDM,
we consider the following $SO(2N \!+\! 1)$ GDM:\\[-2pt]
\beq
\slashed{\cal W}
\stackrel{\mathrm{def}}{=}
G \!
\left[ 
\BA{ccc} 
- 1_{N} & 0 & 0 \\
\\[-8pt]
0 & 1_{N} & 0 \\
\\[-8pt]
0 & 0 & 1
\EA 
\right] \!
G^\dag,
~~
\slashed{\cal W}^{\dag } \!=\! \slashed{\cal W} ,~~
\slashed{\cal W}^2
\!=\!
1_{2N+1}.
\label{SO2N+1GDM}
\eeq\\[-1pt]
In the subsequent section,
we will attempt a different approach to
the derivation of the modified 
$SO(2N \!+\! 1)$
HB EE from
the fermion MF Hamiltonian using
the above form of the GDM
$\slashed{\cal W}$.

\newpage


\setcounter{equation}{0}
\renewcommand{\theequation}{\arabic{section}.\arabic{equation}}

\section{GHB mean-field Hamiltonian and its diagonalization}

~~ 
\def\erw#1{{<\!\!#1\!\!>_g}}
The expectation values (EVs) of the fermion
$SO (2N)$ Lie OPs, i.e.,
the generators of rotation
in $2N$-dimensional Euclidian space,
with respect to the HB WF $| g \!\! >$ are given as\\[-20pt]
\ba
\left.
\BA{ll}
&\erw{E^\alpha_{~\beta }
\!+\!
{\displaystyle \frac{1}{2}\delta_{\alpha \beta }}}
\!=\!
R_{\alpha \beta }
\!=\!
{\displaystyle \frac{1}{2}} \!
\left( \!
\overline{b}_{\alpha i} b_{\beta i}
\!-\!
a^\alpha _{~i} \overline{a}^{\beta }_{~i}
\right)
\!+\!
{\displaystyle \frac{1}{2}}\delta_{\alpha \beta } ,\\
\\[-10pt]
&\erw{E_{\alpha \beta }}
\!=\!
-K_{\alpha \beta }
\!=\!
{\displaystyle \frac{1}{2}} \!
\left( \!
\overline{a}^{\alpha }_{~i} b_{\beta i}
\!-\!
b_{\alpha i} \overline{a}^{ \beta }_{~i}
\right) ,~
\erw{E^{\alpha \beta }}
\!=\!
\overline{K}_{\alpha \beta } , \\
\\[-6pt]
&
\erw{E^{\alpha\gamma } \! E_{\delta\beta }}
\!=\!
R_{\alpha\beta }R_{\gamma\delta }-
R_{\alpha\delta }R_{\gamma\beta }-
\overline{K}_{\alpha\gamma }K_{\delta\beta } .
\EA
\right\}
\label{expectG}
\ea\\[-24pt]

Let the Hamiltonian of fermion system
under consideration be\\[-18pt]
\ba
H
\!=\!
h_{\alpha\beta } \!
\left( \!\! 
E^\alpha_{~\beta } \!+\! \frac{1}{2}\delta_{\alpha\beta } \!\! 
\right)
\!+\!
\frac{1}{4}[\alpha\beta|\gamma\delta]
E^{\alpha\gamma }E_{\delta\beta } ,~~
[\alpha\beta|\gamma\delta]
\!=\!
-
[\alpha\delta|\gamma\beta]
\!=\!
[\gamma\delta|\alpha\beta]
\!=\!
\overline{[\beta\alpha|\delta\gamma]}.
\label{Hamiltonian}
\ea\\[-14pt]
The matrix $h_{\alpha\beta }$ related to a single-particle hamiltonian
includes a chemical
potential and
$[\alpha\beta|\gamma\delta]$
are anti-symmetrized matrix
elements of an interaction potential.
The EV of $H$ with respect to the HB WF $| g \!\! >$
is calculated as\\[-18pt]
\ba
\BA{ll}
\erw H
&\!\!\!=
h_{\alpha\beta }\erw{E^\alpha_{~\beta }
\!\!+\!\!
{\displaystyle \frac{1}{2}}\delta_{\alpha\beta }}
\!\!+\!\!
{\displaystyle \frac{1}{2}}[\alpha\beta|\gamma\delta] \!
\left\{ \!\!
\erw{E^\alpha_{~\beta }
\!+\!
{\displaystyle \frac{1}{2}\delta_{\alpha\beta }}}
\erw{E^\gamma_{~\delta }
\!\!+\!\!
{\displaystyle \frac{1}{2}}\delta_{\gamma\delta }}
\!\!+\!\!
{\displaystyle \frac{1}{2}}\erw{E^{\alpha\gamma }}
\erw{E_{\delta\beta }} \!\!
\right\} \\
\\[-14pt]
&\!\!\!=
h_{\alpha\beta}R_{\alpha\beta}
+
{\displaystyle \frac{1}{2}}[\alpha\beta|\gamma\delta] \!
\left( \!
R_{\alpha\beta}R_{\gamma\delta}
-
{\displaystyle \frac{1}{2}}\overline{K}_{\alpha\gamma}K_{\delta\beta} \!
\right) .
\EA
\label{HexpectG}
\ea\\[-12pt]
The SCF parameters $F \!\!=\!\! (F_{\alpha\beta}) \!\!=\!\! F^\dagger $ and
$D \!\!=\!\! (D_{\alpha\beta}) \!\!=\!\! -D^{\mbox{\scriptsize T}}$
are given by functional derivatives as\\[-18pt]
\ba
\BA{c}
F_{\alpha\beta}
\equiv
{\displaystyle \frac{\partial\erw H}{\partial R_{\alpha\beta}}}
=
h_{\alpha\beta}
+
[\alpha\beta|\gamma\delta] R_{\gamma\delta},~~
D_{\alpha\beta}
\equiv
{\displaystyle \frac{\partial\erw H }{\partial \overline{K}_{\alpha\beta}}}
=
{\displaystyle \frac{1}{2}}[\alpha\gamma|\beta\delta]
(-
K_{\delta\gamma}) .
\EA
\label{SCFparameters}
\ea\\[-16pt]
Instead of Hamiltonian
(\ref{Hamiltonian}),
we introduce the generalized Hartree-Bogoliubov (GHB)
mean-field Hamiltonian (MFH)
$H_{SO(2N+1)}$
with another kind of SCF parameters
$M^{\mbox{\scriptsize T}}
\!=\!
[M_1,\cdots, M_{N}] ,
$\\[-20pt]
\begin{eqnarray}
\begin{array}{rl}
&\!\!\!\!
H_{SO(2N+1)}
=
F_{\alpha \beta} \!
\left( \!
c^{\dag }_\alpha c_\beta
{\displaystyle -\frac{1}{2}\delta_{\alpha \beta}} \!
\right) \!
{\displaystyle 
+\frac{1}{2}
D_{\alpha \beta} c_\alpha c_\beta
-\frac{1}{2}
\overline{D}_{\alpha \beta} c^{\dag }_\alpha c^{\dag }_\beta
}
+
M_\alpha  c^{\dag }_\alpha + \overline{M}_\alpha c_\alpha \\
\\[-6pt]
\!\!
=\!
&\!\!
{\displaystyle \frac{1}{2}} \!
\left[c,~c^{\dag },~{\displaystyle \frac{1}{\sqrt{2}}}\right] \!
\left[ \!
\begin{array}{ccc}
-\overline{F} &  -\overline{D} & \sqrt{2} ~\! \overline{M} \\
\\[-10pt]
D  & F   & \sqrt{2}M \\
\\[-10pt]
\sqrt{2}M^{\mbox{\scriptsize T}} &\sqrt{2}M^{\dag } & 0 
\end{array} \!
\right] \!\!
\left[ \!\!
\begin{array}{c}
c^{\dag } \! , \\
\\[-16pt]
c, \\
\\[-16pt]
{\displaystyle \frac{1}{\sqrt{2}}}
\end{array} \!\!
\right] 
\!=\!
{\displaystyle \frac{1}{2}} \!
\left[c,~c^{\dag },{\displaystyle \frac{1}{\sqrt{2}}}\right] \!\!
\stackrel{\circ}{\mathcal{F}}_{\!\!_{0}} \!\!
\left[ \!\!
\begin{array}{c}
c^{\dag } \! , \\
\\[-16pt]
c, \\
\\[-16pt]
{\displaystyle \frac{1}{\sqrt{2}}}
\end{array} \!\!
\right] .
\end{array}
\label{mean-field Hamiltonian}
\end{eqnarray}\\[-10pt]
We also have studied the GHB-MFH
linear in the Jacobi generator for boson system 
\cite{NishiProvi.18,BerceanuM.92,BerceanuG.92}.
Here,
we introduce
traceless matrices 
$\stackrel{\circ}{\mathcal{F}}$ and
$\mathcal{F}$
expressed in the following forms: \\[-16pt]
\begin{eqnarray}
\left.
\begin{array}{cc}
&
\stackrel{\circ}{\mathcal{F}}
~\!\!\!\!\equiv\!\!
\left[ \!
\begin{array}{ccc}
F    &  D   & \sqrt{2}M \\
\\[-10pt]
-\overline{D} & -\overline{F} & \sqrt{2}~\!  \overline{M} \\
\\[-10pt]
\sqrt{2}M^{\dag } &\sqrt{2}M^{\mbox{\scriptsize T}} & 0 
\end{array} \!
\right] ,~
\stackrel{\circ}{\mathcal{F}}
\equiv\!\!
\left[ \!
\begin{array}{ccc}
0& 1_N& 0 \\
\\[-10pt]
1_N& 0& 0 \\
\\[-10pt]
0& 0& 1 
\end{array}
\right] \!\!
\stackrel{\circ}{\mathcal{F}}_{\!\!_{0}} \!\!
\left[
\begin{array}{ccc}
0& 1_N& 0 \\
\\[-10pt]
1_N& 0& 0 \\
\\[-10pt]
0& 0& 1  
\end{array} \!
\right] , \\
\\[-4pt]
&
\left[ 
\begin{array}{ccc}
0& -1_{N}& 0 \\
\\[-10pt]
-1_{N}& 0& 0 \\
\\[-10pt]
0& 0& 1 
\end{array} 
\right] 
\overline{\stackrel{\circ}{\mathcal{F}}} 
\left[ 
\begin{array}{ccc}
0& -1_{N}& 0 \\
\\[-10pt]
-1_{N}& 0& 0 \\
\\[-10pt]
0& 0& 1  
\end{array} 
\right] \!
=
-
\stackrel{\circ}{\mathcal{F}} ,
\end{array}
\right\}
\label{tildeF}
\end{eqnarray}\\[-12pt]
and\\[-20pt]
\begin{eqnarray}
\mathcal{F} 
\!\equiv\!
\left[ \!
\begin{array}{cc}
F &~\!\! D \\
\\[-2pt]
-\overline{D} &~\!\! -\overline{F} \! 
\end{array} 
\right] .
\label{calF}
\end{eqnarray}
Using the generalized Bogoliubov TR
(\ref{UGdetG}) 
and the OP identity
$z^2 \!-\! \rho^2 \!=\! 1$, 
we can diagonalize the MFH, $H_{SO(2N+1)}$
(\ref{mean-field Hamiltonian})
as follows:\\[-24pt]
\begin{eqnarray}
\begin{array}{rl}
&\!\!\!\!\!\!\!\!\!\!
H_{SO(2N\!+\!1)}
\!\!=\!\!
{\displaystyle \frac{1}{2}} \!\!
\left[d,~d^{\dag },{\displaystyle \frac{1}{\sqrt{2}}}\right] \!\!
\left[
G^\dag \!\!
\left(z \!+\! \rho \right) \!
\stackrel{\circ}{\mathcal{F}}_{\!\!_{0}} \!
\left(z \!-\! \rho \right) \!
G
\right] \!\!
\left[ \!\!
\begin{array}{c}
d^{\dag }, \! \\
\\[-16pt]
d, \\
\\[-16pt]
{\displaystyle \frac{1}{\sqrt{2}}}
\end{array} \!\!
\right] 
\!\!=\!\!
{\displaystyle \frac{1}{2}} \!\!
\left[d,~d^{\dag },{\displaystyle \frac{1}{\sqrt{2}}}\right] \!
G^\dag \!\! \stackrel{\circ}{\mathcal{F}}_{\!\!_{0}} \! G \!\!
\left[ \!\!
\begin{array}{c}
d^{\dag } \! , \\
\\[-16pt]
d, \\
\\[-16pt]
{\displaystyle \frac{1}{\sqrt{2}}}
\end{array} \!\!
\right] \! ,
\end{array} 
\label{diaHm}
\end{eqnarray}
\vspace{-0.7cm}
\begin{eqnarray}
G^\dag
\!\! \stackrel{\circ}{\mathcal{F}}_{\!\!_{0}} \!
G
\!=\!
\left[ \!
\begin{array}{cc}
E_{2N} \!\cdot\! 1_{2N} & 0 \\
\\[-6pt]
0  & 0 
\end{array} \!
\right]
\!\equiv\!
\tilde{E} , ~
E_{2N}
\!=\!
\mbox{diag.} \!
\left(E_1, \cdots, E_N, - E_1, \cdots, - E_N \right) ,
\label{diagonalHm}
\end{eqnarray}
and
$\!
H_{\!SO(2N\!+\!1)}
\!\!=\!\!
\sum_{i=1}^N \!
E_i \!
\left( \!\!
d^{\dag }_i d_i 
\!\!-\!\!
{\displaystyle \frac{1}{2}} \!\!
\right) \!\!
E_i 
\!$
: quasi-particle ({\bf QP}) energy.
Using
(\ref{diagonalHm}) and 
(\ref{tildeF}),
we obtain
\begin{eqnarray}
\begin{array}{l}
\stackrel{\circ}{\mathcal{F}} \!\!
\left[ \!
\begin{array}{ccc}
0& 1_N& 0 \\
\\[-10pt]
1_N& 0& 0 \\
\\[-10pt]
0& 0& 1 
\end{array}
\right] \!
G 
=
\left[ \!
\begin{array}{ccc}
0& 1_N& 0 \\
\\[-10pt]
1_N& 0& 0 \\
\\[-10pt]
0& 0& 1 
\end{array}
\right] \!
G 
\tilde{E} ,
\end{array}
\label{FG}
\end{eqnarray}
which
is explicitly written by
\begin{eqnarray}
\!\!\!\!
\left[ \!
\begin{array}{ccc}
F    & D   & \sqrt{2}M \\
\\
\\[-12pt]
-\overline{D} & -\overline{F} & \sqrt{2}~\!\overline{M} \\
\\
\\[-12pt]
\sqrt{2}M^{\dag } & \sqrt{2}M^{\mbox{\scriptsize T}} &  0 
\end{array} \!
\right] \!
\left[ \!
\begin{array}{ccc}
b
\!+\!
{\displaystyle \frac{1}{1\!+\!z}}{\displaystyle \frac{x}{\sqrt{2}}}
{\displaystyle \frac{y}{\sqrt{2}}} &
\overline{a}
\!-\!
{\displaystyle \frac{1}{1\!+\!z}}{\displaystyle \frac{x}{\sqrt{2}}}
{\displaystyle \frac{\overline{y}}{\sqrt{2}}} &
 {\displaystyle \frac{x}{\sqrt{2}}} \\
\\[-12pt]
a
\!-\!
{\displaystyle \frac{1}{1\!+\!z}} {\displaystyle \frac{\overline{x}}{\sqrt{2}}}
 {\displaystyle \frac{y}{\sqrt{2}}}&
\overline{b}
 \!+\!
{\displaystyle \frac{1}{1\!+\!z}}{\displaystyle \frac{\overline{x}}{\sqrt{2}}}
{\displaystyle \frac{\overline{y}}{\sqrt{2}}} &
- {\displaystyle \frac{\overline{x}}{\sqrt{2}}} \\
\\[-14pt]
{\displaystyle \frac{y}{\sqrt{2}}} & -{\displaystyle \frac{\overline{y}}{\sqrt{2}}} &
z 
\end{array} \!
\right] \nonumber \\ \nonumber
\\[-2pt]
\!=\!
\left[ \!
\begin{array}{ccc}
b
\!+\!
{\displaystyle \frac{1}{1\!+\!z}}{\displaystyle \frac{x}{\sqrt{2}}}
{\displaystyle \frac{y}{\sqrt{2}}} &
\overline{a}
\!-\!
{\displaystyle \frac{1}{1\!+\!z}}{\displaystyle \frac{x}{\sqrt{2}}}
{\displaystyle \frac{\overline{y}}{\sqrt{2}}} &
 {\displaystyle \frac{x}{\sqrt{2}}} \\
\\[-12pt]
a
\!-\!
{\displaystyle \frac{1}{1\!+\!z}} {\displaystyle \frac{\overline{x}}{\sqrt{2}}}
 {\displaystyle \frac{y}{\sqrt{2}}}&
\overline{b}
\!+\!
{\displaystyle \frac{1}{1\!+\!z}}{\displaystyle \frac{\overline{x}}{\sqrt{2}}}
{\displaystyle \frac{\overline{y}}{\sqrt{2}}} &
 -{\displaystyle \frac{\overline{x}}{\sqrt{2}}} \\
\\[-14pt]
{\displaystyle \frac{y}{\sqrt{2}}} & -{\displaystyle \frac{\overline{y}}{\sqrt{2}}} &
z 
\end{array} \!
\right] \!
\left[ \!
\begin{array}{ccc}
e + \varepsilon & 0 & 0 \\
\\
\\[-10pt]
0 & - e - \varepsilon & 0 \\
\\
\\[-10pt]
0 & 0 & 0 
\end{array} \!
\right] ,
\label{extendedeigeneq}
\end{eqnarray}
from the first column in both sides of equations of which,
we get  the following set of equations:
\begin{eqnarray}
\left.
\begin{array}{l}
F b \!+\! D a
\!+\!
{\displaystyle \frac{1}{1\!+\!z}}
\left( \!
F {\displaystyle \frac{x}{\sqrt{2}}}
\!-\!
D {\displaystyle \frac{\overline{x}}{\sqrt{2}}} \!
\right) \!
{\displaystyle \frac{y}{\sqrt{2}}}
\!+\!
\sqrt{2} M
 {\displaystyle \frac{y}{\sqrt{2}}}
 \!=\!
b e
\!+\!
b \varepsilon
\!+\!
{\displaystyle \frac{1}{1\!+\!z}}{\displaystyle \frac{x}{\sqrt{2}}}
{\displaystyle \frac{y}{\sqrt{2}}} \!
\left( e + \varepsilon  \right), \\
\\[-6pt]
-\overline{D} b \!-\! \overline{F} a
\!-\!
{\displaystyle \frac{1}{1\!+\!z}}
\left( \!
\overline{D} {\displaystyle \frac{x}{\sqrt{2}}}
\!-\!
\overline{F} {\displaystyle \frac{\overline{x}}{\sqrt{2}}} \!
\right) \!
{\displaystyle \frac{y}{\sqrt{2}}}
\!+\!
\sqrt{2} ~\! \overline{M}
{\displaystyle \frac{y}{\sqrt{2}}}
\!=\!
a e
+
a \varepsilon
\!-\!
{\displaystyle \frac{1}{1\!+\!z}}{\displaystyle \frac{\overline{x}}{\sqrt{2}}}
{\displaystyle \frac{y}{\sqrt{2}}} \!
\left( e + \varepsilon  \right), \\
\\[-10pt]
\sqrt{2}\!M^{\dag } b \!+\! \sqrt{2}\!M^{\mbox{\scriptsize T}} a
\!+\!
{\displaystyle \frac{1}{1\!+\!z}} \!
\left( \!
\sqrt{2}\!M^{\dag } {\displaystyle \frac{x}{\sqrt{2}}}
\!-\!
\sqrt{2}\!M^{\mbox{\scriptsize T}}
{\displaystyle \frac{\overline{x}}{\sqrt{2}}} \!
\right) \!
{\displaystyle \frac{y}{\sqrt{2}}}
\!=\!
{\displaystyle \frac{y}{\sqrt{2}}} e
\!+\!
{\displaystyle \frac{y}{\sqrt{2}}} \varepsilon .
\end{array}
\right\}
\label{eigeneq00}
\end{eqnarray}\\[-10pt]
Then we have\\[-20pt]
\begin{eqnarray}
\!\!\!\!
\left.
\begin{array}{c}
F b \!+\! D a = b e , ~~~~
\overline{D} b \!+\! \overline{F} a = - a e ,  \\
\\[-6pt]
{\displaystyle \frac{1}{1\!+\!z}}
\left( \!
F {\displaystyle \frac{x}{\sqrt{2}}}
\!-\!
D {\displaystyle \frac{\overline{x}}{\sqrt{2}}} \!
\right) \!
{\displaystyle \frac{y}{\sqrt{2}}}
\!+\!
\sqrt{2} M
 {\displaystyle \frac{y}{\sqrt{2}}}
\!=\!
b \varepsilon
\!+\!
{\displaystyle \frac{1}{1\!+\!z}}{\displaystyle \frac{x}{\sqrt{2}}}
{\displaystyle \frac{y}{\sqrt{2}}} \!
\left( e + \varepsilon  \right) , \\
\\[-6pt]
{\displaystyle \frac{1}{1\!+\!z}}
\left( \!
\overline{D} {\displaystyle \frac{x}{\sqrt{2}}}
\!-\!
\overline{F} {\displaystyle \frac{\overline{x}}{\sqrt{2}}} \!
\right) \!
{\displaystyle \frac{y}{\sqrt{2}}}
\!-\!
\sqrt{2} ~\! \overline{M}
{\displaystyle \frac{y}{\sqrt{2}}}
\!=\! 
- a \varepsilon
\!+\!
{\displaystyle \frac{1}{1\!+\!z}}{\displaystyle \frac{\overline{x}}{\sqrt{2}}}
{\displaystyle \frac{y}{\sqrt{2}}} \!
\left( e + \varepsilon  \right) ,  \\
\\[-10pt]
\sqrt{2}\! M^{\dag } b \!+\! \sqrt{2}\! M^{\mbox{\scriptsize T}} a
\!=\! 
{\displaystyle \frac{y}{\sqrt{2}}}
e , ~~~~
{\displaystyle \frac{1}{1\!+\!z}} \!
\left( \!\!
\sqrt{2}\!M^{\dag } {\displaystyle \frac{x}{\sqrt{2}}}
\!-\!
\sqrt{2}\!M^{\mbox{\scriptsize T}}
{\displaystyle \frac{\overline{x}}{\sqrt{2}}} \!
\right) 
\!=\! 
\varepsilon ,
\end{array} \!\!
\right\}
\label{eigeneq}
\end{eqnarray}
By through the second column
in both sides of equations of
(\ref{extendedeigeneq}),
we get the following set of equations:\\[-20pt]
\begin{eqnarray}
\left.
\begin{array}{l}
F \overline{a} \!+\! D \overline{b}
\!-\!
{\displaystyle \frac{1}{1\!+\!z}} \!
\left( \!
F {\displaystyle \frac{x}{\sqrt{2}}}
\!-\!
D {\displaystyle \frac{\overline{x}}{\sqrt{2}}} \!
\right) \!
{\displaystyle \frac{\overline{y}}{\sqrt{2}}}
\!-\!
\sqrt{2} M
{\displaystyle \frac{\overline{y}}{\sqrt{2}}} 
\!=\!
- \overline{a} e
-
\overline{a} \varepsilon
\!+\!
{\displaystyle \frac{1}{1\!+\!z}}{\displaystyle \frac{x}{\sqrt{2}}}
{\displaystyle \frac{\overline{y}}{\sqrt{2}}}
\left( e + \varepsilon  \right) , \\
\\[-6pt]
- \overline{D} \overline{a} \!-\! \overline{F} \overline{b}
\!+\!
{\displaystyle \frac{1}{1\!+\!z}}
\left( \!
\overline{D} {\displaystyle \frac{x}{\sqrt{2}}}
\!-\!
\overline{F} {\displaystyle \frac{\overline{x}}{\sqrt{2}}} \!
\right) \!
{\displaystyle \frac{\overline{y}}{\sqrt{2}}}
\!-\!
\sqrt{2} ~\! \overline{M}
{\displaystyle \frac{\overline{y}}{\sqrt{2}}}
\!=\!
-
\overline{b} e
\!-\!
\overline{b} \varepsilon
\!-\!
{\displaystyle \frac{1}{1\!+\!z}}{\displaystyle \frac{\overline{x}}{\sqrt{2}}}
{\displaystyle \frac{\overline{y}}{\sqrt{2}}}
\left( e + \varepsilon  \right) , \\
\\[-6pt]
\sqrt{2}\!M^{\dag } \overline{a}
\!+\!
\sqrt{2}\!M^{\mbox{\scriptsize T}} \overline{b}
\!-\!
{\displaystyle \frac{1}{1\!+\!z}}
\left( \!
\sqrt{2}\!M^{\dag } {\displaystyle \frac{x}{\sqrt{2}}}
\!-\!
\sqrt{2}\!M^{\mbox{\scriptsize T}}
{\displaystyle \frac{\overline{x}}{\sqrt{2}}} \!
\right) \!
{\displaystyle \frac{\overline{y}}{\sqrt{2}}}
\!=\!
{\displaystyle \frac{\overline{y}}{\sqrt{2}}} e
\!+\!
{\displaystyle \frac{\overline{y}}{\sqrt{2}}} \varepsilon ,
\end{array}
\right\}
\label{eigeneq20}
\end{eqnarray}\\[-10pt]
from which we have\\[-20pt]
\begin{eqnarray}
\!\!\!\!
\left.
\begin{array}{c}
F \overline{a} \!+\! D \overline{b}  = - \overline{a} e , ~~~~
\overline{D} \overline{a} \!+\! \overline{F} \overline{b} = \overline{b} e , \\
\\[-6pt]
{\displaystyle \frac{1}{1\!+\!z}} \!
\left( \!
F {\displaystyle \frac{x}{\sqrt{2}}}
\!-\!
D {\displaystyle \frac{\overline{x}}{\sqrt{2}}} \!
\right) \!
{\displaystyle \frac{\overline{y}}{\sqrt{2}}}
\!+\!
\sqrt{2} M
 {\displaystyle \frac{\overline{y}}{\sqrt{2}}}
\!=\! 
\overline{a} \varepsilon
\!-\!
{\displaystyle \frac{1}{1\!+\!z}}{\displaystyle \frac{x}{\sqrt{2}}}
{\displaystyle \frac{\overline{y}}{\sqrt{2}}}
\left( e + \varepsilon  \right) ,  \\
\\[-6pt]
{\displaystyle \frac{1}{1\!+\!z}}
\left( \!
\overline{D} {\displaystyle \frac{x}{\sqrt{2}}}
\!-\!
\overline{F} {\displaystyle \frac{\overline{x}}{\sqrt{2}}} \!
\right) \!
{\displaystyle \frac{\overline{y}}{\sqrt{2}}}
\!-\!
\sqrt{2} ~\! \overline{M}
{\displaystyle \frac{\overline{y}}{\sqrt{2}}}
\!=\!
- \overline{b} \varepsilon
\!-\!
{\displaystyle \frac{1}{1\!+\!z}}{\displaystyle \frac{\overline{x}}{\sqrt{2}}}
{\displaystyle \frac{\overline{y}}{\sqrt{2}}}
\left( e + \varepsilon  \right) , \\
\\[-6pt]
\sqrt{2}\!M^{\dag } \overline{a}
\!+\!
\sqrt{2}\!M^{\mbox{\scriptsize T}} \overline{b}
\!=\! 
{\displaystyle \frac{\overline{y}}{\sqrt{2}}}
e , ~~~~
{\displaystyle \frac{1}{1\!+\!z}} \!
\left( \!\!
\sqrt{2}\!M^{\dag } {\displaystyle \frac{x}{\sqrt{2}}}
\!-\!
\sqrt{2}\!M^{\mbox{\scriptsize T}}
{\displaystyle \frac{\overline{x}}{\sqrt{2}}} \!
\right) \!
\!=\!
- \varepsilon .
\end{array} \!\!
\right\}
\label{eigeneq2}
\end{eqnarray}\\[-10pt]
Finally, by through the third column
in both sides of equations of
(\ref{extendedeigeneq}),
we get equations\\[-14pt]
\begin{eqnarray}
\!\!\!\!
\left.
\begin{array}{c}
F {\displaystyle \frac{x}{\sqrt{2}}} 
\!-\!
D {\displaystyle \frac{\overline{x}}{\sqrt{2}}}
\!+\!
\sqrt{2} z M 
\!=\!
0 ,~~
- \overline{D} {\displaystyle \frac{x}{\sqrt{2}}} 
\!+\!
\overline{F} \! {\displaystyle \frac{\overline{x}}{\sqrt{2}}}
\!+\!
\sqrt{2} z \overline{M} 
\!=\!
0 , \\
\\[-10pt]
\sqrt{2}M^{\dag } {\displaystyle \frac{x}{\sqrt{2}}} 
\!-\! 
\sqrt{2}M^{\mbox{\scriptsize T}} {\displaystyle \frac{\overline{x}}{\sqrt{2}}}
\!=\!
0 .
\end{array}
\right\}
\label{eigeneq3}
\end{eqnarray}
The first and second equations of
(\ref{eigeneq}) and (\ref{eigeneq2})
are set of the well-known fermion Bogoliubov equations
with eigenvalue $e$. The division of $E$ into $e$ and $\varepsilon$
is a useful manner to treat the MF eigenvalue equation (EE)
and to make clear its structure
through each step of calculative ways.
The additional eigenvalue, $\varepsilon$, however,
turns out to be  $\varepsilon \!=\! 0$
due to the relations
(\ref{eigeneq3}).  
 
From now on,
we adopt the method developed in the GHB-MFT for boson system
\cite{NishiProvi.18}.
Multiplying the first of 
(\ref{eigeneq}) 
by $b^{\dag}$ 
and 
that of
(\ref{eigeneq2})
by $a^{\mbox{\scriptsize T}}$ 
from the right, 
we obtain
\begin{eqnarray}
F b b^{\dag} \!+\! D a b^{\dag}
\!=\!
b e b^{\dag}, ~~
F \overline{a} a^{\mbox{\scriptsize T}}
\!+\!
D \overline{b} ~\!\! a^{\mbox{\scriptsize T}}
\!=\! 
-\overline{a} e a^{\mbox{\scriptsize T}} .
\label{eigeneq4}
\end{eqnarray}
Also multiplying the first of 
(\ref{eigeneq}) 
by $a^{\dag}$
and 
that of
(\ref{eigeneq2})
by $b^{\mbox{\scriptsize T}}$ 
from the right,
we obtain
\begin{eqnarray}
F b a^{\dag} \!+\! D a a^{\dag}
\!=\!
b e a^{\dag}, ~~
F \overline{a} b^{\mbox{\scriptsize T}}
\!+\!
D \overline{b} b^{\mbox{\scriptsize T}}
\!=\! 
-\overline{a} e b^{\mbox{\scriptsize T}} .
\label{eigeneq5}
\end{eqnarray}
Adding second of
(\ref{eigeneq4}) 
to first
and that of 
(\ref{eigeneq4}) 
to first
and using
(\ref{RepBogotrans}),
we have
\begin{eqnarray}
F
\!=\!
-
\left(
\overline{a} e a^{\mbox{\scriptsize T}} - b e b^{ \dag}
\right)
\!=\!
F^{\dag}, ~~
D
\!=\!
-
\left(
\overline{a} e b^{\mbox{\scriptsize T}} - b e a^{\dag}
\right)
\!=\!
- D^{\mbox{\scriptsize T}}.
\label{FD}
\end{eqnarray}
Multiplying the third equation of 
(\ref{eigeneq})
by $b^{\dag }$ and that of
(\ref{eigeneq20})
by $a^{\mbox{\scriptsize T}}$ from the right, 
we have\\[-14pt] 
\begin{eqnarray}
\left.
\begin{array}{rl}
&\!\!
\sqrt{2} z M^{\dag } b b^{\dag }
\!+\!
\sqrt{2} z M^{\mbox{\scriptsize T}} a b^{\dag }
\!=\!
\left( \!  
{\displaystyle \frac{x^{\mbox{\scriptsize T}} }{\sqrt{2}}} \! a 
\!-\!
{\displaystyle \frac{x^{\dag }}{\sqrt{2}}} b \!
\right) \!
e b^{\dag }
\!-\!
\left(1 - z \right) \!
{\displaystyle \frac{y }{\sqrt{2}}} e b^{\dag }, \\
\\[-4pt]
&\!\!
\sqrt{2} z M^{\dag} \overline{a} a^{\mbox{\scriptsize T}}
\!+\!
\sqrt{2} z M^{\mbox{\scriptsize T} } \overline{b} a^{\mbox{\scriptsize T}}
\!=\!
\left( \!   
{\displaystyle \frac{x^{\dag }}{\sqrt{2}}} \overline{a} 
\!-\!
{\displaystyle \frac{x^{\mbox{\scriptsize T}} }{\sqrt{2}}}\overline{b} \!
\right) \!
e a^{\mbox{\scriptsize T}}
\!-\!
\left(1 - z \right) \!
{\displaystyle \frac{\overline{y} }{\sqrt{2}}} e a^{\mbox{\scriptsize T}} ,
\end{array}
\right\} 
\label{eigeneq6}
\end{eqnarray}
where we have used the relation
$
zy
\!=\!
x^{\mbox{\scriptsize T}} A
\!-\!
x^{\dag } B
\!=\!
x^{\mbox{\scriptsize T}} a
\!-\!
x^{\dag } b
\!-\!
\left(1 - z \right) y .
$
In
(\ref{eigeneq6}),
adding the equation in the second line to
the equation in the first line
and using
(\ref{RepBogotrans}),
we get\\[-16pt]
\ba
\sqrt{2} z M^{\dag}
\!=\!
-
{\displaystyle \frac{x^{\dag }}{\sqrt{2}}}
F
-
{\displaystyle \frac{x^{\mbox{\scriptsize T}} }{\sqrt{2}}}
\overline{D}
-
\left(1 - z \right)
\left( \!
{\displaystyle \frac{y }{\sqrt{2}}} e b^{\dag }
\!+\!
{\displaystyle \frac{\overline{y} }{\sqrt{2}}} e a^{\mbox{\scriptsize T}} \!
\right) ,
\label{eigeneq7}
\ea\\[-12pt]
where we have used the relations
(\ref{FD}).
The form of equation
(\ref{eigeneq7})
is changed as\\[-16pt]
\ba
\left.
\begin{array}{l}
\sqrt{2} z M
\!=\!
- F {\displaystyle \frac{x}{\sqrt{2}}} 
\!+\!
D {\displaystyle \frac{\overline{x}}{\sqrt{2}}}
\!-\!
\left(1 - z \right)
\left( \!
b e {\displaystyle \frac{y^\dag }{\sqrt{2}}} 
\!+\!
\overline{a} e {\displaystyle \frac{y^{\mbox{\scriptsize T}}}{\sqrt{2}}} \!
\right) , \\
\\[-6pt]
\sqrt{2} z \overline{M}
\!=\!
- \overline{F} {\displaystyle \frac{\overline{x}}{\sqrt{2}}} 
\!+\!
\overline{D} {\displaystyle \frac{x}{\sqrt{2}}}
\!-\!
\left(1 - z \right)
\left( \!
\overline{b} e {\displaystyle \frac{y^{\mbox{\scriptsize T}} }{\sqrt{2}}} 
\!+\!
a e {\displaystyle \frac{y^{\dag}}{\sqrt{2}}} \!
\right) ,
\end{array}
\right\}
\label{eigeneq8}
\ea\\[-8pt]
owing to
(\ref{eigeneq3}), 
which gives the additional conditions\\[-20pt]
\ba
b e y^\dag  
\!+\!
\overline{a} e y^{\mbox{\scriptsize T}}
\!=\!
0,~~
\overline{b} e y^{\mbox{\scriptsize T}} 
\!+\!
a e y^{\dag}
\!=\!
0 .
\label{eigeneqcondition}
\ea\\[-18pt]
In
(\ref{eigeneq8}),
multiplying the first equation in the first line by $F$ and
the second one in the second line by $D$ and
adding them,
we get\\[-18pt] 
\begin{eqnarray}
\BA{ll}
F \sqrt{2} z M
\!+\!
D \sqrt{2} z \overline{M}
\!\!&\!=\!
- F F {\displaystyle \frac{x}{\sqrt{2}}} 
\!+\!
F D {\displaystyle \frac{\overline{x}}{\sqrt{2}}}
\!-\!
\left(1 - z \right)
\left( \!
F b e {\displaystyle \frac{y^\dag }{\sqrt{2}}} 
\!+\!
F \overline{a} e {\displaystyle \frac{y^{\mbox{\scriptsize T}}}{\sqrt{2}}} \!
\right) \\
\\[-10pt]
&~~\!-\!
D \overline{F} {\displaystyle \frac{\overline{x}}{\sqrt{2}}} 
\!+\!
D \overline{D} {\displaystyle \frac{x}{\sqrt{2}}}
\!-\!
\left(1 - z \right)
\left( \!
D \overline{b} e {\displaystyle \frac{y^{\mbox{\scriptsize T}} }{\sqrt{2}}} 
\!+\!
D a e {\displaystyle \frac{y^{\dag}}{\sqrt{2}}} \!
\right)   \\
\\[-10pt]
&\!=\!
-
\left( FF \!\!-\!\! D \overline{D} \right) \!
{\displaystyle \frac{x}{\sqrt{2}}}
\!+\!
\left( FD \!\!-\!\! D \overline{F} \right) \!
{\displaystyle \frac{\overline{x}}{\sqrt{2}}} 
\!-\!
\left(1 \!-\! z \right) \!
\left( \!
b e e {\displaystyle \frac{y^\dag }{\sqrt{2}}} 
\!-\!
\overline{a} e e {\displaystyle \frac{y^{\mbox{\scriptsize T}}}{\sqrt{2}}} \!
\right) .
\EA
\label{FzMDzM}
\end{eqnarray}\\[-6pt]
If the conditions
$ FD \!-\! D \overline{F} \!=\! 0$ and
$b e e y^\dag 
\!-\!
\overline{a} e e y^{\mbox{\scriptsize T}}
\!=\! 0$,
which lead the condition
$e b^{-1} \overline{a} \!=\! - b^{\!-1} \overline{a} e$,
are satisfied,
then we have the expression for $x$  and $x^\dag$ as\\[-14pt]
\ba
\left.
\BA{ll}
{\displaystyle \frac{x}{\sqrt{2}}}
&
\!=\!
-
\left( F F^\dag \!+\! D D^\dag \right)^{-1} \!
\left( 
F \sqrt{2} z M
\!+\!
D \sqrt{2} z \overline{M} 
\right) , \\
\\[-6pt]
{\displaystyle \frac{x^\dag}{\sqrt{2}}}
&
\!=\!
-
\left( 
\sqrt{2} z M^\dag F^\dag
\!+\!
\sqrt{2} z M^{\mbox{\scriptsize T}} D^\dag
\right) \!
\left( F F^\dag \!+\! D D^\dag \right)^{-1}  .
\EA
\right\}
\label{solutionx}
\ea\\[-24pt]

On the other hand,
combining the equations in the first and the second lines of
(\ref{eigeneq00}) and (\ref{eigeneq})
and those in the first line of
(\ref{eigeneq3})
and using
$\varepsilon \!=\! 0$,
we have the following relations:\\[-14pt]
\begin{eqnarray}
\left.
\begin{array}{ll}
&
\left( \!
1 \!-\! {\displaystyle \frac{z}{1\!+\!z}} \!
\right) \!
\sqrt{2} M 
{\displaystyle \frac{y}{\sqrt{2}}} 
 \!=\!
{\displaystyle \frac{1}{1\!+\!z}}{\displaystyle \frac{x}{\sqrt{2}}}
{\displaystyle \frac{y}{\sqrt{2}}}
e
\!+\!
\left( \!
b 
\!+\!   
{\displaystyle \frac{1}{1\!+\!z}}{\displaystyle \frac{x}{\sqrt{2}}}
{\displaystyle \frac{y}{\sqrt{2}}} \!
\right)
\varepsilon
\!=\!
{\displaystyle \frac{1}{1\!+\!z}}{\displaystyle \frac{x}{\sqrt{2}}}
{\displaystyle \frac{y}{\sqrt{2}}}
e ,  \\
\\[2pt]
&
\left( \!
1 \!-\! {\displaystyle \frac{z}{1\!+\!z}} \!
\right) \!
\sqrt{2} ~\! \overline{M} 
{\displaystyle \frac{y}{\sqrt{2}}}
\!=\!
-
{\displaystyle \frac{1}{1\!+\!z}}{\displaystyle \frac{\overline{x}}{\sqrt{2}}}
{\displaystyle \frac{y}{\sqrt{2}}}
e , 
\end{array} \!\!
\right\}
\label{eigeneq00xy}
\end{eqnarray}\\[-8pt]
and further combining the equations in the first and the second lines of
(\ref{eigeneq20}) and (\ref{eigeneq2})
and those in the first line of
(\ref{eigeneq3})
and using
$\varepsilon \!=\! 0$,
we have the following relations:\\[-14pt]
\begin{eqnarray}
\begin{array}{l}
\sqrt{2} M \!
{\displaystyle \frac{\overline{y}}{\sqrt{2}}}
\!=\!
-
{\displaystyle \frac{x}{\sqrt{2}}}
{\displaystyle \frac{\overline{y}}{\sqrt{2}}}
e , ~~
\sqrt{2} ~\! \overline{M} \!
{\displaystyle \frac{\overline{y}}{\sqrt{2}}}
\!=\!
{\displaystyle \frac{\overline{x}}{\sqrt{2}}}
{\displaystyle \frac{\overline{y}}{\sqrt{2}}}
e .
\end{array}
\label{eigeneq00xbary}
\end{eqnarray}\\[-12pt]
The relations
(\ref{eigeneqcondition})
and
$e b^{-1} \overline{a} \!=\! - b^{\!-1} \overline{a} e$
give the consistent result
with
(\ref{eigeneq8})
as\\[-16pt]
\ba
b e e {\displaystyle \frac{y^\dag }{\sqrt{2}}} 
\!-\!
\overline{a} e e {\displaystyle \frac{y^{\mbox{\scriptsize T}}}{\sqrt{2}}}
\!=\!
b e b^{-1} \!
\left( \!
b e {\displaystyle \frac{y^\dag }{\sqrt{2}}} 
\!+\!
\overline{a} e {\displaystyle \frac{y^{\mbox{\scriptsize T}}}{\sqrt{2}}} \!
\right)
\!=\!
0 .
\label{consitentcondition}
\ea\\[-20pt]

In
(\ref{solutionx}),
multiplying the equations in the first and second lines by
${\displaystyle \frac{y}{\sqrt{2}}}$
and
${\displaystyle \frac{y^\dag}{\sqrt{2}}}$
from the right and left, respectively,
we obtain the equations\\[-14pt]
\ba
\left.
\BA{ll}
{\displaystyle \frac{x}{\sqrt{2}}}
{\displaystyle \frac{y}{\sqrt{2}}}
&
\!=\!
-
\left( F F^\dag \!+\! D D^\dag \right)^{-1} \!
\left( \!
F \sqrt{2} z M
{\displaystyle \frac{y}{\sqrt{2}}}
\!+\!
D \sqrt{2} z \overline{M}
{\displaystyle \frac{y}{\sqrt{2}}} \!
\right) \\
\\[-12pt]
&
\!=\!
-
\left( F F^\dag \!\!+\!\! D D^\dag \right)^{-1} \!
z \!
\left( \!
F {\displaystyle \frac{x}{\sqrt{2}}}
\!-\!
D {\displaystyle \frac{\overline{x}}{\sqrt{2}}} \!
\right) \!\!
{\displaystyle \frac{y}{\sqrt{2}}}
e
\!=\!
\left( F F^\dag \!+\! D D^\dag \right)^{\!-1} \!
z 
\sqrt{2} z M
{\displaystyle \frac{y}{\sqrt{2}}}
e , \\
\\[-8pt]
{\displaystyle \frac{y^\dag}{\sqrt{2}}}
{\displaystyle \frac{x^\dag}{\sqrt{2}}}
&
\!=\!
-
\left( \!
{\displaystyle \frac{y^\dag}{\sqrt{2}}} 
\sqrt{2} z M^\dag F^\dag
\!+\!
{\displaystyle \frac{y^\dag}{\sqrt{2}}}
\sqrt{2} z M^{\mbox{\scriptsize T}} D^\dag \!
\right) \!
\left( F F^\dag \!+\! D D^\dag \right)^{\!-1} \\
\\[-12pt]
&
\!\!=\!\!
-
z 
e
{\displaystyle \frac{y^\dag}{\sqrt{2}}} \!
\left( \!
{\displaystyle \frac{x^\dag}{\sqrt{2}}} F^\dag
\!-\!
{\displaystyle \frac{x^{\mbox{\scriptsize T}}}{\sqrt{2}}} D^\dag \!
\right) \!\!
\left( F F^\dag \!+\! D D^\dag \right)^{\!-1}
\!=\!
z
e
{\displaystyle \frac{y^\dag}{\sqrt{2}}}
\sqrt{2} z M^\dag \!
\left( F F^\dag \!+\! D D^\dag \right)^{\!-1} \! ,
\EA \!\!
\right\}
\label{solutionxy}
\ea\\[-4pt]
where we have used the relations
(\ref{eigeneq00xy})
and
(\ref{eigeneq00xbary}).
Further multiplying the first and second equations in
(\ref{solutionxy}) by
${\displaystyle \frac{y^\dag}{\sqrt{2}}}$
and
${\displaystyle \frac{y}{\sqrt{2}}}$
from the right and left, respectively,
we have\\[-8pt]
\ba
\left.
\BA{ll}
{\displaystyle \frac{x}{\sqrt{2}}}
{\displaystyle \frac{y}{\sqrt{2}}}
{\displaystyle \frac{y^\dag}{\sqrt{2}}}
&
=
{\displaystyle \frac{x}{\sqrt{2}}}
{\displaystyle \frac{1 - z^2 }{2}} 
=
\left( F F^\dag + D D^\dag \right)^{-1} 
z 
\sqrt{2} z M
{\displaystyle \frac{y}{\sqrt{2}}}
e
{\displaystyle \frac{y^\dag}{\sqrt{2}}} , \\
\\[-4pt]
{\displaystyle \frac{y}{\sqrt{2}}}
{\displaystyle \frac{y^\dag}{\sqrt{2}}}
{\displaystyle \frac{x^\dag}{\sqrt{2}}}
&
=
{\displaystyle \frac{1 - z^2 }{2}}
{\displaystyle \frac{x^\dag}{\sqrt{2}}} 
= 
{\displaystyle \frac{y}{\sqrt{2}}}
e
{\displaystyle \frac{y^\dag}{\sqrt{2}}}
z
\sqrt{2} z M^\dag 
\left( F F^\dag + D D^\dag \right)^{-1}  .
\EA \!\!
\right\}
\label{solutionxyydag}
\ea\\[-6pt]
Then, at last we reach the following expressions for the vectors
${\displaystyle \frac{x}{\sqrt{2}}}$
and
${\displaystyle \frac{x^\dag}{\sqrt{2}}}$:\\[-12pt]
\ba
\left.
\BA{ll}
{\displaystyle \frac{x}{\sqrt{2}}}
&
\!\!\!=\!
\left( F F^\dag \!+\! D D^\dag \right)^{-1} \!
{\displaystyle \frac{2z^2}{1 - z^2}}  
\sqrt{2} M
{\displaystyle \frac{y}{\sqrt{2}}}
e
{\displaystyle \frac{y^\dag}{\sqrt{2}}} 
=
{\displaystyle \frac{2z^2}{1 - z^2}}  
<\!e\!> \!
\left( F F^\dag \!+\! D D^\dag \right)^{-1} \!
\sqrt{2} M, \\
\\[-8pt]
{\displaystyle \frac{x^\dag}{\sqrt{2}}}
&
\!\!\!=\!
{\displaystyle \frac{2z^2}{1 - z^2}}
{\displaystyle \frac{y}{\sqrt{2}}}
e
{\displaystyle \frac{y^\dag}{\sqrt{2}}}
\sqrt{2} M^\dag \!
\left( F F^\dag \!+\! D D^\dag \right)^{-1} \!
=
{\displaystyle \frac{2z^2}{1 - z^2}}  
<\!e\!> \!
\sqrt{2} M^\dag \!
\left( F F^\dag \!+\! D D^\dag \right)^{\!-1} \!\! ,
\EA \!\!
\right\}
\label{finalsolutionx}
\ea\\[-8pt]
where $<\!e\!>$ is defined as
$
<\!e\!> 
\equiv
{\displaystyle \frac{y}{\sqrt{2}}}
e
{\displaystyle \frac{y^\dag}{\sqrt{2}}} 
$.
Therefore, the
$<\!e\!>$
stands for the averaged value of all the eigenvalue distribution.
Thus, this is the first time that
the final solutions for
the vectors
${\displaystyle \frac{x}{\sqrt{2}}}$
and
${\displaystyle \frac{x^\dag}{\sqrt{2}}}$
could be derived within the present
framework of the $SO(2N\!+\!1)$ MFT.
Such a situation takes place also for the Jacobi-algebra 
MFT for a boson system
with $N(2N\!+\!1)\!+\!2$ Jacobi generators
\cite{NishiProvi.18}.
The inner product of the above vectors
is shown to
lead us to the relation\\[-14pt]
\ba
{\displaystyle \frac{x^\dag}{\sqrt{2}}}
{\displaystyle \frac{x}{\sqrt{2}}}
=
{\displaystyle \frac{1 - z^2 }{2}}
=
{\displaystyle \frac{4z^4}{(1 - z^2)^2}}  
~ 2
<\!e\!>^2 \!
M^\dag 
\left( F F^\dag \!+\! D D^\dag \right)^{-2} 
M ,
\label{xdagx}
\ea\\[-8pt]
which is an appreciably interesting result in the $SO(2N\!+\!1)$
MFT
and is simply rewritten as\\[-12pt]
\ba
{\displaystyle \frac{16z^4}{(1 - z^2)^3 }}
<\!e\!>^2 \!
M^\dag 
\left( F F^\dag \!+\! D D^\dag \right)^{-2} 
M
=
1 ,
\label{MdagM}
\ea\\[-8pt]
which designates that the magnitude of
the additional SCF parameter $M$
is inevitably restricted by the behavior of
the SCF parameters $F$ and $D$
governed by the condition
$FD \!-\! D \overline{F} \!=\! 0$.
Remember that this condition is one of
the crucial condition in
(\ref{FzMDzM})
to derive the equations
for vectors
$\!{\displaystyle \frac{x}{\sqrt{2}}}\!$
and
$\!{\displaystyle \frac{x^\dag}{\sqrt{2}}}\!$
(\ref{solutionx})
which reflect the special aspect of
the $SO(2N\!\!+\!\!1)$ MFT.
Such a result should not be a surprised consequence that 
the relation
(\ref{MdagM})
is very similar to the relation obtained
in the GHB-MFT for a boson system
\cite{NishiProvi.18}.
This is
because we have adopted the same manner 
of mathematical computation as the manner
that is done for the boson case.


\newpage

\setcounter{equation}{0}
\renewcommand{\theequation}{\arabic{section}.\arabic{equation}}

\section{Mean-field approach using another form of GDM}

According to equations
(\ref{chiraloptrans}) and (\ref{relAtoaXY}),
we introduce a matrix $g_x$ represented by\\[-6pt] 
\beq
g_x
\!\!=\!\!\!
\left[ \!\!\!
\BA{cc} 
1_{N} \!-\! {\displaystyle \frac{1}{\sqrt{1 \!\!+\!\! z}}}
{\displaystyle \frac{\overline{x}}{\sqrt{2}}}
{\displaystyle \frac{1}{\sqrt{1 \!\!+\!\! z}}}
{\displaystyle \frac{x^{\mbox{\scriptsize T}}}{\sqrt{2}}}&
{\displaystyle \frac{1}{\sqrt{1 \!\!+\!\! z}}}
{\displaystyle \frac{\overline{x}}{\sqrt{2}}}
{\displaystyle \frac{1}{\sqrt{1 \!\!+\!\! z}}}
{\displaystyle \frac{x^\dag}{\sqrt{2}}} \\
\\[-6pt]
{\displaystyle \frac{1}{\sqrt{1 \!\!+\!\! z}}}
{\displaystyle \frac{x}{\sqrt{2}}}
{\displaystyle \frac{1}{\sqrt{1 \!\!+\!\! z}}}
{\displaystyle \frac{x^{\mbox{\scriptsize T}}}{\sqrt{2}}}&
\!\!\!\!
1_{N} \!-\! {\displaystyle \frac{1}{\sqrt{1 \!\!+\!\! z}}}
{\displaystyle \frac{x}{\sqrt{2}}}
{\displaystyle \frac{1}{\sqrt{1 \!\!+\!\! z}}}
{\displaystyle \frac{x^\dag}{\sqrt{2}}}
\EA \!\!\!
\right] ,~
g^\dag_x
\!=\!
g_x .
\label{gx}
\eeq  
Then,
using the another form of GDM,
$\slashed{W}\!$
(\ref{anotherGDM}),
the explicit expression for $\slashed{\cal W}$
is given as\\[-14pt] 
\beqa
\BA{ll}
\slashed{\cal W} 
&\!\!\!=\!
\left[ \!\!
\BA{ccc}
&
&\!\!
{\displaystyle -\frac{\overline{x}}{\sqrt{2}}} \\ 
&\!\!g_x&
\\[-8pt]
&
&\!\!
{\displaystyle ~~\frac{x}{\sqrt{2}}} \\
{\displaystyle \frac{x^{\mbox{\scriptsize T}}}{\sqrt{2}}} &\!\! 
{\displaystyle -\frac{x^\dag}{\sqrt{2}}} &\!\! z 
\EA \!\!
\right] \!\!
\left[ \!
\BA{ccc}
&
&
0 \\ \\[-6pt]
&\slashed{W}&
\\ \\[-6pt]
&
&
0 \\ \\[-6pt]
0 & 
0 & 1
\EA \!
\right] \!\!
\left[ \!\!
\BA{ccc}
&
&\!\!
{\displaystyle ~~\frac{\overline{x}}{\sqrt{2}}} \\[-2pt]
&\!\!g^\dag_x&
\\[-8pt]
&
&\!\!
-{\displaystyle \frac{x}{\sqrt{2}}} \\
{\displaystyle -\frac{x^{\mbox{\scriptsize T}}}{\sqrt{2}}} &\!\!
{\displaystyle \frac{x^\dag}{\sqrt{2}}} &\!\! z 
\EA \!\!
\right] \\
\\[-12pt] 
&
\!\!\!=\!
\left[ \!\!\!
\BA{ccc}
&
&\!\!
{\displaystyle -\frac{\overline{x}}{\sqrt{2}}} \\[-2pt] 
&\!\!g_x&
\\[-8pt]
&
&\!\!
{\displaystyle ~~\frac{x}{\sqrt{2}}} \\
{\displaystyle \frac{x^{\mbox{\scriptsize T}}}{\sqrt{2}}} &\!\! 
{\displaystyle -\frac{x^\dag}{\sqrt{2}}} &\!\! z 
\EA \!\!
\right] \!\!
\left[ \!\!\!\!
\BA{cccc}
&
~~~~~~~
\slashed{W} \! g^\dag_x
&\!\!\!\!\!\!\!\!\!\!\!
\BA{c}
~~~~~~~~~
\slashed{W}
\EA
&\!\!\!\!\!\!\!\!
\left[
\BA{c}
{\displaystyle \frac{\overline{x}}{\sqrt{2}}}\\
\\[-2pt]
-{\displaystyle \frac{x}{\sqrt{2}}}
\EA
\right] \\[-2pt] 
\\[-8pt]
&\!\!\!\!\!\!\!\!\!\!\!\!\!\!\!
\left[{\displaystyle -\frac{x^{\mbox{\scriptsize T}}}{\sqrt{2}}}\right.
&\!\!\!\!\!\!\!\!\!
\left.{\displaystyle \frac{x^\dag}{\sqrt{2}}} \right] 
&\!\!\!\!\!\! z
\EA \!\!\!
\right] \\
\\[-12pt] 
&
\!\!\!=\!
\left[ \!\!\!\!\!\!\!\!
\BA{cccc}
&
~~~~~~~
g_x \slashed{W} \! g^\dag_x
&\!\!\!\!\!\!\!\!\!\!\!
\BA{c}
~~~~~~~~~
g_x \slashed{W}
\EA
&\!\!\!\!\!\!\!\!
\left[
\BA{c}
{\displaystyle \frac{\overline{x}}{\sqrt{2}}}\\
\\[-2pt]
-{\displaystyle \frac{x}{\sqrt{2}}}
\EA
\right] \\[-2pt] 
\\[-8pt]
&\!\!\!\!\!\!\!\!\!\!\!\!\!\!\!\!\!
\left[{\displaystyle \frac{x^{\mbox{\scriptsize T}}}{\sqrt{2}}}\right.
&\!\!\!\!\!\!\!\!\!\!\!\!\!\!\!\!\!\!\!\!\!\!\!
\left.{\displaystyle -\frac{x^\dag}{\sqrt{2}}} \right] \!\!
\slashed{W} \! g^\dag_x
&\!\!\!\!\!\! z^2
\EA \!\!\!
\right]
+
\left[ \!\!
\BA{ccc} 
{\displaystyle \frac{\overline{x}}{\sqrt{2}}}
{\displaystyle \frac{x^{\mbox{\scriptsize T}}}{\sqrt{2}}} & 0 &
{\displaystyle -z\frac{\overline{x}}{\sqrt{2}}} \\
\\[-10pt]
0 & {\displaystyle \frac{x}{\sqrt{2}}}
{\displaystyle \frac{x^\dag}{\sqrt{2}}} &
{\displaystyle ~~z\frac{x}{\sqrt{2}}} \\
\\[-10pt]
{\displaystyle -z\frac{x^{\mbox{\scriptsize T}}}{\sqrt{2}}} &
{\displaystyle z\frac{x^\dag}{\sqrt{2}}}& 0
\EA \!\!
\right]  ,
\EA 
\label{SO2N+1GDM2}
\eeqa\\[-10pt]
where
$g_x \slashed{W} g^\dag_x, ~g_x \slashed{W}$
and
$\slashed{W} g^\dag_x$
are given by\\[-14pt] 
\beq
~~g_x \slashed{W} g^\dag_x
\!=\!
\left[ \!
\BA{cc} 
2 \rho - 1_{N} &
- 2 \overline{\kappa} \\
\\[-8pt]
2 \kappa &
- 2 \overline{\rho} + 1_{N} 
\EA \!
\right] ,~
\BA{c}
\rho
=
R \!-\! \overline{L}
{\displaystyle \frac{1}{\sqrt{1 \!\!+\!\! z}}}
{\displaystyle \frac{x^{\mbox{\scriptsize T}}}{\sqrt{2}}}
\!-\!
{\displaystyle \frac{1}{\sqrt{1 \!\!+\!\! z}}}
{\displaystyle \frac{\overline{x}}{\sqrt{2}}}
L^{\mbox{\scriptsize T}}, \\
\\[-14pt]
\kappa
=
K  \!-\! L
{\displaystyle \frac{1}{\sqrt{1 \!\!+\!\! z}}}
{\displaystyle \frac{x^{\mbox{\scriptsize T}}}{\sqrt{2}}}
\!+\! 
{\displaystyle \frac{1}{\sqrt{1 \!\!+\!\! z}}}
{\displaystyle \frac{x}{\sqrt{2}}}
L^{\mbox{\scriptsize T}},
\EA
\label{gxWgxdag}
\eeq
\vspace{-0.4cm}  
\beqa
\BA{ll}
&g_x \slashed{W} 
=
\left[ \!
\BA{cc}
2 R
-
1_{N}
-
{\displaystyle \frac{2 \sqrt{2}}{\sqrt{1 \!\!+\!\! z}}}
{\displaystyle \frac{\overline{x}}{\sqrt{2}}}
{\displaystyle \frac{L^{\mbox{\scriptsize T}}}{\sqrt{2}}}&
- 2 \overline{K}
-
{\displaystyle \frac{2 \sqrt{2}}{\sqrt{1 \!\!+\!\! z}}}
{\displaystyle \frac{\overline{x}}{\sqrt{2}}}
{\displaystyle \frac{L^\dag}{\sqrt{2}}} \\
\\[-8pt]
2 K
+
{\displaystyle \frac{2 \sqrt{2}}{\sqrt{1 \!\!+\!\! z}}}
{\displaystyle \frac{x}{\sqrt{2}}}
{\displaystyle \frac{L^{\mbox{\scriptsize T}}}{\sqrt{2}}}&
\!\!\!\!
- 2 \overline{R}
+
1_{N}
+
{\displaystyle \frac{2 \sqrt{2}}{\sqrt{1 \!\!+\!\! z}}}
{\displaystyle \frac{x}{\sqrt{2}}}
{\displaystyle \frac{L^\dag}{\sqrt{2}}}
\EA \!
\right] , \\
\\[-6pt]
&\slashed{W} \! g^\dag_x
=
\left[ \!
\BA{cc} 
2 R
-
1_{N}
-
{\displaystyle \frac{2 \sqrt{2}}{\sqrt{1 \!\!+\!\! z}}}
{\displaystyle \frac{\overline{L}}{\sqrt{2}}}
{\displaystyle \frac{x^{\mbox{\scriptsize T}}}{\sqrt{2}}}&
- 2 \overline{K}
+
{\displaystyle \frac{2 \sqrt{2}}{\sqrt{1 \!\!+\!\! z}}}
{\displaystyle \frac{\overline{L}}{\sqrt{2}}}
{\displaystyle \frac{x^\dag}{\sqrt{2}}} \\
\\[-8pt]
2 K
-
{\displaystyle \frac{2 \sqrt{2}}{\sqrt{1 \!\!+\!\! z}}}
{\displaystyle \frac{L}{\sqrt{2}}}
{\displaystyle \frac{x^{\mbox{\scriptsize T}}}{\sqrt{2}}}&
\!\!\!\!
- 2 \overline{R}
+
1_{N}
+
{\displaystyle \frac{2 \sqrt{2}}{\sqrt{1 \!\!+\!\! z}}}
{\displaystyle \frac{L}{\sqrt{2}}}
{\displaystyle \frac{x^\dag}{\sqrt{2}}}
\EA \!
\right] , 
\EA
\label{gxWgx}
\eeqa
\vspace{-0.6cm}
\beqa
\BA{l}
\!\!\!\!\!\!\!\!\!\!\!\!\!\!\!\!\!\!\!\!\!\!\!\!\!\!\!\!\!\!
\!\!\!\!\!\!\!\!\!\!\!\!\!\!\!\!\!\!\!\!\!\!\!\!\!\!\!\!\!\!
L
=
{\displaystyle \frac{1}{\sqrt{2}}}
{\displaystyle \frac{1}{\sqrt{1 + z}}} 
\left( 
\overline{R} x 
+ 
K \overline{x}
-
{\displaystyle \frac{1}{2}}
x 
\right)  .
\EA
 \label{gxWgx2}
 \eeqa
 The column and row vectors in
 the first matrix in the last line of
 (\ref{SO2N+1GDM2})
 are calculated as
\beqa
\BA{ll}
&
g_x \slashed{W} \!
\left[ \!\!
\BA{c} 
~~{\displaystyle \frac{\overline{x}}{\sqrt{2}}}\\
\\[-4pt]
-{\displaystyle \frac{x}{\sqrt{2}}}
\EA \!\!
\right] \!
\!=\!
\left[ \!
\BA{cc}
2 R
-
1_{N}
-
{\displaystyle \frac{2 \sqrt{2}}{\sqrt{1 \!\!+\!\! z}}}
{\displaystyle \frac{\overline{x}}{\sqrt{2}}}
{\displaystyle \frac{L^{\mbox{\scriptsize T}}}{\sqrt{2}}}&
- 2 \overline{K}
-
{\displaystyle \frac{2 \sqrt{2}}{\sqrt{1 \!\!+\!\! z}}}
{\displaystyle \frac{\overline{x}}{\sqrt{2}}}
{\displaystyle \frac{L^\dag}{\sqrt{2}}} \\
\\[-6pt]
2 K
+
{\displaystyle \frac{2 \sqrt{2}}{\sqrt{1 \!\!+\!\! z}}}
{\displaystyle \frac{x}{\sqrt{2}}}
{\displaystyle \frac{L^{\mbox{\scriptsize T}}}{\sqrt{2}}}&
\!\!\!\!
- 2 \overline{R}
+
1_{N}
+
{\displaystyle \frac{2 \sqrt{2}}{\sqrt{1 \!\!+\!\! z}}}
{\displaystyle \frac{x}{\sqrt{2}}}
{\displaystyle \frac{L^\dag}{\sqrt{2}}}
\EA \!
\right] \!\!
\left[ \!\!
\BA{c} 
~~{\displaystyle \frac{\overline{x}}{\sqrt{2}}}\\
\\[-2pt]
-{\displaystyle \frac{x}{\sqrt{2}}}
\EA \!\!
\right] \\
\\
&
\!\!=\!
\left[ \!\!\!
\BA{c} 
\sqrt{2} \!
\left( \!
R \overline{x} \!+\! \overline{K} x
\!-\!
{\displaystyle \frac{1}{2}}
\overline{x} \!
\right)
\!-\!
{\displaystyle \frac{1}{\sqrt{1 + z}}}
\left( 
\overline{x} L^{\mbox{\scriptsize T}} \overline{x}
\!-\!
\overline{x} L^{\dag} x 
\right) \\
\\
\sqrt{2} \!
\left( \!
\overline{R} x \!+\! K \overline{x}
\!-\!
{\displaystyle \frac{1}{2}}
x \!
\right)
\!-\!
{\displaystyle \frac{1}{\sqrt{1 + z}}}
\left( 
x L^{\dag} x
\!-\!
x L^{\mbox{\scriptsize T}} \overline{x} 
\right)
\EA \!\!\!
\right] 
\!\!=\!\!
\left[ \!\!\!
\BA{c} 
2 \sqrt{2} \sqrt{1 \!+\! z} ~\! {\displaystyle \frac{\overline{L}}{\sqrt{2}}} \\
\\[-2pt]
2 \sqrt{2} \sqrt{1 \!+\! z} ~\! {\displaystyle \frac{L}{\sqrt{2}}}
\EA \!\!\!
\right] 
\!\!=\!
\slashed{W} \!\!
\left[ \!\!
\BA{c} 
~~{\displaystyle \frac{\overline{x}}{\sqrt{2}}}\\
\\[-2pt]
-{\displaystyle \frac{x}{\sqrt{2}}}
\EA \!\!
\right] \! ,
\EA
\label{gxWx}
\eeqa
\vspace{-0.5cm}
\beqa
\BA{ll}
&
\left[  
{\displaystyle \frac{x^{\mbox{\scriptsize T}}}{\sqrt{2}}}~~~
- \! {\displaystyle \frac{x^\dag}{\sqrt{2}}} 
\right] \!
\slashed{W} \! g^\dag_x 
\!=\!
\left[  
{\displaystyle \frac{x^{\mbox{\scriptsize T}}}{\sqrt{2}}}~~~
- \! {\displaystyle \frac{x^\dag}{\sqrt{2}}} 
\right] \!\!
\left[ \!\!
\BA{cc}
2 R
\!-\!
1_{N}
\!-
{\displaystyle \frac{2 \sqrt{2}}{\sqrt{1 \!\!+\!\! z}}}
{\displaystyle \frac{\overline{L}}{\sqrt{2}}}
{\displaystyle \frac{x^{\mbox{\scriptsize T}}}{\sqrt{2}}}&
- 2 \overline{K}
\!-
{\displaystyle \frac{2 \sqrt{2}}{\sqrt{1 \!\!+\!\! z}}}
{\displaystyle \frac{\overline{L}}{\sqrt{2}}}
{\displaystyle \frac{x^\dag}{\sqrt{2}}} \\
\\[-4pt]
2 K
\!+\!
{\displaystyle \frac{2 \sqrt{2}}{\sqrt{1 \!\!+\!\! z}}}
{\displaystyle \frac{L}{\sqrt{2}}}
{\displaystyle \frac{x^{\mbox{\scriptsize T}}}{\sqrt{2}}}&
\!\!\!\!
- 2 \overline{R}
\!+\!
1_{N}
\!+\!
{\displaystyle \frac{2 \sqrt{2}}{\sqrt{1 \!\!+\!\! z}}}
{\displaystyle \frac{L}{\sqrt{2}}}
{\displaystyle \frac{x^\dag}{\sqrt{2}}}
\EA \!\!
\right]  \\
\\[-6pt]
&
\!\!=\!
\left[ \! 
\sqrt{2} 
\left( 
x^{\mbox{\scriptsize T}} \! R \!-\! x^{\dag} K
\!-\!
{\displaystyle \frac{1}{2}}
x^{\!\mbox{\scriptsize T}} 
\right)
\!-\!
{\displaystyle \frac{1}{\sqrt{1 \!+\! z}}}
\left( 
x^{\!\mbox{\scriptsize T}} \overline{L} x^{\!\mbox{\scriptsize T}} 
\!-\!
x^{\dag} L x^{\!\mbox{\scriptsize T}} 
\right)
\right.\\
\\[-6pt]
&
\left.
~~~~~~~~~~~~~~~~~~~~~~~~~~~~~~~~~~~~~~~~
\sqrt{2} 
\left( 
x^{\dag} \overline{R} \!-\! x^{\!\mbox{\scriptsize T}} \overline{K} 
\!-\!
{\displaystyle \frac{1}{2}} 
x^{\dag} 
\right)
\!-\!
{\displaystyle \frac{1}{\sqrt{1 \!+\! z}}} 
\left( 
x^{\dag} L x^{\dag}
\!-\!
x^{\!\mbox{\scriptsize T}} \overline{L} x^\dag 
\right) 
\right] \\
\\[-6pt]
&
\!\!=\!
\left[ 
2 \sqrt{2} \sqrt{1 \!+\! z} ~\! 
{\displaystyle \frac{L^{\mbox{\scriptsize T}}}{\sqrt{2}}}
~~~~~
2 \sqrt{2} \sqrt{1 \!+\! z} ~\! 
{\displaystyle \frac{L^\dag}{\sqrt{2}}} 
\right]
\!=\!
\left[ 
{\displaystyle \frac{x^{\mbox{\scriptsize T}}}{\sqrt{2}}}~~~
-{\displaystyle \frac{x^\dag}{\sqrt{2}}} 
\right] 
\slashed{W} ,
\EA
\label{xWgx}
\eeqa
where we have used
$
x L^{\dag} x
\!-\!
x L^{\mbox{\scriptsize T}} \overline{x}
=
0
$
and
$
x^{\dag} L x^{\dag}
\!-\!
x^{\!\mbox{\scriptsize T}} \overline{L} x^\dag
=
0
$.
Thus we reach our desired goal to find the final expression for 
the $SO(2N \!+\! 1)$ GDM
$\slashed{\cal W}$ as
\beq
\slashed{\cal W}
\!=\!
\slashed{\cal W}_{\!\rho ~\!\! \kappa}
\!+\!
\slashed{\cal W}_{\!x} 
\!=\!\!
\left[ \!\!
\BA{ccc}
2 \rho  \!-\! 1_{N} & -2 \overline{\kappa} &
2 \! \sqrt{\!1 \!\!+\!\! z} ~\! \overline{L} \\
\\[14pt]
2 \kappa & - 2 \overline{\rho} \!+\! 1_{N} &
2 \! \sqrt{\!1 \!\!+\!\! z} ~\!\! L \\
\\[14pt]
2 \! \sqrt{\!1 \!\!+\!\! z} ~\!\! L^{\!\mbox{\scriptsize T}} & 
2 \! \sqrt{\!1 \!\!+\!\! z} ~\!\! L^\dag  & z^2 
\EA \!\!\!
\right] 
\!+\!
\left[ \!\!\!
\BA{ccc} 
{\displaystyle \frac{\overline{x}}{\sqrt{2}}}
{\displaystyle \frac{x^{\mbox{\scriptsize T}}}{\sqrt{2}}} & 0 &
{\displaystyle -z\frac{\overline{x}}{\sqrt{2}}} \\
\\[-6pt]
0 & {\displaystyle \frac{x}{\sqrt{2}}}
{\displaystyle \frac{x^\dag}{\sqrt{2}}} &
{\displaystyle ~~z\frac{x}{\sqrt{2}}} \\
\\[-6pt]
{\displaystyle -z\frac{x^{\mbox{\scriptsize T}}}{\sqrt{2}}} &
{\displaystyle z\frac{x^\dag}{\sqrt{2}}}& 0
\EA \!\!\!
\right] .
\label{finalW}
\eeq

Parallel to the way made in
(\ref{SO2N+1GDM2}),  
the GHB MF OP ${\cal F}$,
which is formally of the same form as the first equation of
(\ref{tildeF}),
is transformed to $\slashed{\cal F}$ as
\beqa
\BA{l}
\slashed{\cal F} 
\!=\!
\left[ \!\!
\BA{ccc}
&
&\!\!
{\displaystyle -\frac{\overline{x}}{\sqrt{2}}} \\ 
&\!\!g_x&
\\[-6pt]
&
&\!\!
{\displaystyle ~~\frac{x}{\sqrt{2}}} \\
{\displaystyle \frac{x^{\mbox{\scriptsize T}}}{\sqrt{2}}} &\!\! 
{\displaystyle -\frac{x^\dag}{\sqrt{2}}} &\!\! z 
\EA \!\!
\right] \!\!
\left[ \!\!
\BA{ccc}
&
&
\sqrt{2} M \\ \\[-6pt]
&\!\!\!\! \widehat{\cal F}&
\\ \\[-6pt]
&
&
\sqrt{2} ~ \! \overline{M} \\ \\[-6pt]
\sqrt{2} M^\dag & 
\sqrt{2} M^{\mbox{\scriptsize T}} & 0
\EA \!\! 
\right] \!\!
\left[ \!\!
\BA{ccc}
&
&\!\!
{\displaystyle ~~\frac{\overline{x}}{\sqrt{2}}} \\
&\!\!g^\dag_x&
\\[-8pt]
&
&\!\!
-{\displaystyle \frac{x}{\sqrt{2}}} \\
{\displaystyle -\frac{x^{\mbox{\scriptsize T}}}{\sqrt{2}}} &\!\!
{\displaystyle \frac{x^\dag}{\sqrt{2}}} &\!\! z 
\EA \!\!
\right] ,
\EA
\label{transF}
\eeqa
in which,
instead of the matrix ${\cal F}$
(\ref{calF}),
here we use 
a matrix $\widehat{\cal F}$
which modifys ${\cal F}$ as
\beq
\widehat{\cal F}
=
\left[ \!
\BA{cc} 
\widehat{F}&  \widehat{D} \\
\\[-4pt]
- \overline{\widehat{D}} & - \overline{\widehat{F}} 
\EA \!
\right] ,~
\BA{c}
\widehat{F}_{\alpha\beta}
=
h_{\alpha\beta}
+
[\alpha\beta|\gamma\delta] \rho_{\gamma\delta} , \\
\\[-6pt]
\widehat{D}_{\alpha\beta}
=
{\displaystyle \frac{1}{2}}[\alpha\gamma|\beta\delta]
\left( -\kappa_{\delta\gamma} \right) .
\EA
\label{Fmat}
\eeq
The transformed MF OP
$\slashed{\cal F}$
(\ref{transF})
is rewritten as\\[-8pt]
\beqa
\!\!\!\!\!\!\!\!
\BA{ll}
&
\slashed{\cal F}
\!=\!
\left[ \!\!\!
\BA{ccc}
&
&\!\!
{\displaystyle -\frac{\overline{x}}{\sqrt{2}}} \\ 
&\!\!g_x&
\\[-6pt]
&
&\!\!
{\displaystyle ~~\frac{x}{\sqrt{2}}} \\
{\displaystyle \frac{x^{\mbox{\scriptsize T}}}{\sqrt{2}}} &\!\! 
{\displaystyle -\frac{x^\dag}{\sqrt{2}}} &\!\! z 
\EA \!\!
\right] \!\!
\left[ \!\!
\BA{ccc}
&\!\!\!\!
\widehat{{\cal F}} g^\dag_x~
\!+\!
\left[ \!\!
\BA{cc} 
-M \! x^{\mbox{\scriptsize T}} & M \! x^\dag \\
\\[8pt]
-\overline{M} \! x^{\mbox{\scriptsize T}} & \overline{M} \! x^\dag \!
\EA \!\!
\right]
&
\widehat{{\cal F}} \!
\left[ \!\!\!
\BA{c}
{\displaystyle \frac{\overline{x}}{\sqrt{\! 2}}} \\
\\[-6pt]
{\displaystyle-\frac{x}{\sqrt{\! 2}}}
\EA \!\!\!
\right] \!
\!+\!
z \!
\left[ \!\!
\BA{c}
\sqrt{2} M \\
\\[12pt]
\sqrt{2} ~\! \overline{M}
\EA \!\!
\right] \\
\\[-4pt]
&\!\!\!\!\!\! 
\left[ 
\sqrt{2} M^\dag
~~~~
\sqrt{2} M^{\mbox{\scriptsize T}}  
\right] 
g^\dag_x
&\!\!\!\!
M^\dag \overline{x}
\!-\!
M^{\mbox{\scriptsize T}} x
\EA \!\!
\right] \\
\\[2pt] 
&
\!=\!\!
\left[ \!\!\!\!\!\!\!\!
\BA{ccc}
&\!\!\!\!
g_x \widehat{{\cal F}} g^\dag_x
\!+\!
g_x \!\!
\left[  \!\!\!
\BA{cc} 
-M \! x^{\!\mbox{\scriptsize T}} &\!\! M \! x^\dag \\
\\[6pt]
-\overline{M} \! x^{\!\mbox{\scriptsize T}} &\!\! \overline{M} \! x^\dag \!
\EA  \!\!\!
\right]
\!\!+\!\!
\left[  \!\!\!
\BA{cc} 
- \overline{x} \! M^{\dag} &\!\! - \overline{x} \! M^{\!\mbox{\scriptsize T}} \\
\\[6pt]
x \! M^{\dag} &\!\! x \! M^{\!\mbox{\scriptsize T}} \!
\EA  \!\!\!\!
\right] \!\!
g^\dag_x
&\!\!\!\!\!\!\!\!
\left[  
g_x \! \widehat{{\cal F}}
\!\!+\!\!
M^{\! \mbox{\scriptsize T}} \! x
\!\!-\!\!
M^{\! \dag} \overline{x}
\right] \!\!
\left[ \!\!\!
\BA{c}
{\displaystyle \frac{\overline{x}}{\sqrt{\! 2}}} \\
\\[-12pt]
{\displaystyle-\frac{x}{\sqrt{\! 2}}}
 \EA \!\!\!
\right] \!
\!+\!\!
z g_x \!\!\!
\left[ \!\!\!
\BA{c}
\sqrt{\! 2} M \\
\\[4pt]
\sqrt{\! 2} \overline{M}
\EA \!\!\!
\right] \\
\\[-4pt]
&
\left[ \!
{\displaystyle \frac{x^{\! \mbox{\scriptsize T}}}{\sqrt{\! 2}}}
~~{\displaystyle-\frac{x^{\! \dag}}{\sqrt{\! 2}}} \!
\right] \!\!
\left[ \! 
\widehat{{\cal F}} g^{\! \dag}_x
\!\!+\!\! 
\left[ \!\!\!
\BA{cc} 
- \! M \! x^{\! \mbox{\scriptsize T}} &\!\! M \! x^{\! \dag} \\
\\[6pt]
- \! \overline{M} \! x^{\! \mbox{\scriptsize T}} &\!\! 
\overline{M} \! x^{\! \dag} \!
\EA \!\!\!
\right] \!
\right] \!
\!\!+\!\!
z \!
\left[ \! \sqrt{\! 2} \! M^{\! \dag} ~~\! 
\sqrt{\! 2} \! M^{\! \mbox{\scriptsize T}} \!
\right] \! g^{\! \dag}_x 
&
\left[ \!
{\displaystyle \frac{x^{\! \mbox{\scriptsize T}}}{\sqrt{\! 2}}}
~~~
{\displaystyle-\frac{x^{\! \dag}}{\sqrt{\! 2}}} \!
\right] \!
\widehat{{\cal F}} \!
\left[ \!\!\!
\BA{c} 
~{\displaystyle \frac{\overline{x}}{\sqrt{\! 2}}} \\
\\[-10pt]
{\displaystyle-\frac{x}{\sqrt{\! 2}}} 
\EA \!\!\!
\right]
\EA \!\!\!
\right] \! .
\EA
\label{transF2}
\eeqa
At a glance,
it is considered that
the OP
$\slashed{\cal F}$
doesn't keep both the properties of the hermitian conjugate,
the characteristic of the $SO(2N \!+\! 1)$ group and
of the traceless matrix, i.e., $\mbox{Tr}\slashed{\cal F} \!\!=\!\! 0$.
However, paying attention to the underlying relations
obtained through some calculations
\beqa
\!\!\!\!\!\!
\left[ \!
{\displaystyle \frac{x^{\! \mbox{\scriptsize T}}}{\sqrt{\! 2}}}
~~~ 
{\displaystyle-\frac{x^{\! \dag}}{\sqrt{\! 2}}} \!
\right] \!\! 
\left[ \!\!\!
\BA{cc} 
- \! M \! x^{\! \mbox{\scriptsize T}} & M \! x^{\! \dag} \\
\\[-8pt]
- \! \overline{M} \! x^{\! \mbox{\scriptsize T}} & \overline{M} \! x^{\! \dag} 
\EA \!\!\!
\right]
\!=\!
{\displaystyle \frac{1}{\sqrt{\! 2}}} \!
\left[ \!
- x^{ \! \mbox{\scriptsize T}} \!\! M \! x^{ \! \mbox{\scriptsize T}} 
\!\!+\!\!
x^{\dag} \! \overline{M} \! x^{\! \mbox{\scriptsize T}}
~~~
x^{ \! \mbox{\scriptsize T}} \!\! M \! x^{\dag}
\!\!-\!\!
x^{\dag} \! \overline{M} \! x^{\dag} \!
\right]
\!\!=\!\!
\left[ \!
{\displaystyle \frac{x^{\! \mbox{\scriptsize T}}}{\sqrt{\! 2}}}  
~~ 
{\displaystyle-\frac{x^{\! \dag}}{\sqrt{\! 2}}} \!
\right] \! 
\left[ \!
x^{\dag} \! \overline{M}
\!\!-\!\!
x^{ \! \mbox{\scriptsize T}} \!\! M
\right] \! ,
\label{xTM}
\eeqa
\vspace{-0.3cm}
\beqa
\!\!\!\!\!\!
\left[ \!
{\displaystyle \frac{x^{\! \mbox{\scriptsize T}}}{\sqrt{\! 2}}}
~~~~
{\displaystyle-\frac{x^{\! \dag}}{\sqrt{\! 2}}} \!
\right] \!\!
\widehat{{\cal F}} \!\!
\left[ \!\!\!
\BA{c} 
~{\displaystyle \frac{\overline{x}}{\sqrt{\! 2}}} \\
\\[-12pt]
{\displaystyle-\frac{x}{\sqrt{\! 2}}} 
\EA \!\!\!
\right] \!
\!=\!
{\displaystyle \frac{1}{2}} \!
\left[ \!
x^{\! \mbox{\scriptsize T}}
~~
- x^{\! \dag} \!
\right] \!\!\!
\left[ \!\!\!
\BA{cc} 
\widehat{F}&  \widehat{D} \\
\\[-6pt]
- \overline{\widehat{D}} & - \overline{\widehat{F}} 
\EA \!\!\!
\right]  \!\!\!
\left[ \!\!\!
\BA{c} 
~\overline{x} \\
\\[-2pt]
- x
\EA \!\!\!
\right] \!
\!=\!
{\displaystyle \frac{1}{2}} \!
\left( \!
x^{\! \mbox{\scriptsize T}} \! \widehat{F} \overline{x}
\!-\!
x^{\! \mbox{\scriptsize T}} \! \widehat{D} x
\!-\!
x^{\! \dag} \overline{\widehat{F}} x
\!+\!
x^{\! \dag} \overline{\widehat{D}} \overline{x} \!
\right)
\!=
0 ,
\label{xdagFx}
\eeqa\\[-10pt]
it turns out that 
the transformed MF OP
$\slashed{\cal F}$
becomes to restore exactly the hermitian conjugate,
the characteristic of the $SO(2N \!+\! 1)$ group and
the tracelessness, $\mbox{Tr}\slashed{\cal F} \! = \! 0$.

In the matrix representation for $\slashed{\cal F}$,
using
(\ref{Fmat})
and
(\ref{gx}),
$g_x \widehat{{\cal F}} \! g^\dag_x$
is calculated as
\beqa
\!\!\!\!
\BA{ll}
g_x \widehat{{\cal F}} \! g^\dag_x
&
\!\!\!=\!
\left[ \!\!\!
\BA{cc} 
1_{N} \!-\! {\displaystyle \frac{1}{1 \!\!+\!\! z}}
{\displaystyle \frac{\overline{x}}{\sqrt{2}}}
{\displaystyle \frac{x^{\mbox{\scriptsize T}}}{\sqrt{2}}}&
{\displaystyle \frac{1}{1 \!\!+\!\! z}}
{\displaystyle \frac{\overline{x}}{\sqrt{2}}}
{\displaystyle \frac{x^\dag}{\sqrt{2}}} \\
\\[-6pt]
{\displaystyle \frac{1}{1 \!\!+\!\! z}}
{\displaystyle \frac{x}{\sqrt{2}}}
{\displaystyle \frac{x^{\mbox{\scriptsize T}}}{\sqrt{2}}}&
\!\!\!\!
1_{N} \!-\! {\displaystyle \frac{1}{1 \!\!+\!\! z}}
{\displaystyle \frac{x}{\sqrt{2}}}
{\displaystyle \frac{x^\dag}{\sqrt{2}}}
\EA \!\!\!
\right] \!\!\!
\left[ \!\!\!
\BA{cc} 
\widehat{F}&  \widehat{D} \\
\\[14pt]
- \overline{\widehat{D}} & - \overline{\widehat{F}} 
\EA \!\!\!
\right] \!\!\! 
\left[ \!\!\!
\BA{cc} 
1_{N} \!-\! {\displaystyle \frac{1}{1 \!\!+\!\! z}}
{\displaystyle \frac{\overline{x}}{\sqrt{2}}}
{\displaystyle \frac{x^{\mbox{\scriptsize T}}}{\sqrt{2}}}&
{\displaystyle \frac{1}{1 \!\!+\!\! z}}
{\displaystyle \frac{\overline{x}}{\sqrt{2}}}
{\displaystyle \frac{x^\dag}{\sqrt{2}}} \\
\\[-6pt]
{\displaystyle \frac{1}{1 \!\!+\!\! z}} 
{\displaystyle \frac{x}{\sqrt{2}}}
{\displaystyle \frac{x^{\mbox{\scriptsize T}}}{\sqrt{2}}}&
\!\!\!\!
1_{N} \!-\! {\displaystyle \frac{1}{1 \!\!+\!\! z}}
{\displaystyle \frac{x}{\sqrt{2}}}
{\displaystyle \frac{x^\dag}{\sqrt{2}}}
\EA \!\!\!
\right] \\
\\
&
\!\equiv
\widehat{\slashed{\cal F}}
=\!
\left[ \!\!
\BA{cc} 
\widehat{\slashed{F}}&  \widehat{\slashed{D}} \\
\\[6pt]
- \overline{\widehat{\slashed{D}}} & - \overline{\widehat{\slashed{F}}} 
\EA \!\!
\right] ,~
\BA{c}
\widehat{\slashed{F}}
=
\widehat{F}
\!-\!
{\displaystyle \frac{1}{1 \!\!+\!\! z}} \!
\left[ \!
{\displaystyle \frac{x}{\sqrt{2}}} \!
\left( \!
{\displaystyle \frac{x^\dag}{\sqrt{2}}} F
 \!+\!
{\displaystyle \frac{x^{\mbox{\scriptsize T}}}{\sqrt{2}}} \overline{D} \!
\right)
\!+\!
\left( \!
F {\displaystyle \frac{x}{\sqrt{2}}}
\!-\!
D {\displaystyle \frac{\overline{x}}{\sqrt{2}}} \!
\right) \!
{\displaystyle \frac{x^\dag}{\sqrt{2}}} \!
\right] , \\
\\[-2pt]
\widehat{\slashed{D}}
=
\widehat{D}
\!-\!
{\displaystyle \frac{1}{1 \!\!+\!\! z}} \!
\left[ \!
{\displaystyle \frac{x}{\sqrt{2}}} \!
\left( \!
{\displaystyle \frac{x^{\mbox{\scriptsize T}}}{\sqrt{2}}} \overline{F} 
 \!+\!
{\displaystyle \frac{x^{\dag}}{\sqrt{2}}} D \!
\right)
\!-\!
\left( \!
F {\displaystyle \frac{x}{\sqrt{2}}}
\!-\!
D {\displaystyle \frac{\overline{x}}{\sqrt{2}}} \!
\right) \!
{\displaystyle \frac{x^{\mbox{\scriptsize T}}}{\sqrt{2}}} \!
\right] ,
\EA
\EA
\label{transbygx}
\eeqa
where we have used the relations derivable with the aid of
(\ref{gxWgxdag})\\[-16pt]
\beqa
\!\!\!\!\!\!\!\!
\BA{ll}
&
\left( \!
{\displaystyle \frac{x^\dag}{\sqrt{2}}} \widehat{F}
 \!+\!
{\displaystyle \frac{x^{\mbox{\scriptsize T}}}{\sqrt{2}}} \overline{\widehat{D}} \!
\right)_{\!\!\beta}
\!=\!
{\displaystyle \frac{\overline{x}_{\alpha}}{\sqrt{2}}} \widehat{F}_{\alpha \beta}
 \!+\!
{\displaystyle \frac{x_{\alpha}}{\sqrt{2}}} \overline{\widehat{D}}_{\alpha \beta} \\
\\[-8pt]
&
\!=\!
{\displaystyle \frac{\overline{x}_{\alpha}}{\sqrt{2}}} \!
\left\{ \!\!
F_{\alpha \beta}
\!-\!
{\displaystyle \frac{1}{\sqrt{1 \!\!+\!\! z}}}
[\alpha\beta|\gamma\delta] \!
\left( \!\!
{\displaystyle \frac{\overline{x}_{\gamma}}{\sqrt{2}}}
L _{\delta}
\!\!+\!\!
\overline{L} _{\gamma}
{\displaystyle \frac{x_{\delta}}{\sqrt{2}}} \!
\right) \!\!
\right\}  
\!+\!
{\displaystyle \frac{x_{\alpha}}{\sqrt{2}}} \!
\left\{ \!\!
\overline{D}_{\alpha \beta}
\!+\!
{\displaystyle \frac{1}{\sqrt{1 \!\!+\!\! z}}}
{\displaystyle \frac{1}{2}}
\overline{[\alpha\gamma|\beta\delta]} \!
\overline{
\left( \!\!
{\displaystyle \frac{x_{\delta}}{\sqrt{2}}} L _{\gamma}
\!-\!
L _{\delta} {\displaystyle \frac{\overline{x}_{\gamma}}{\sqrt{2}}} \!
\right)
} \!\!
\right\}  \\
\\[-8pt]
&
\!=\!
{\displaystyle \frac{\overline{x}_{\alpha}}{\sqrt{2}}} F_{\alpha \beta}
 \!+\!
{\displaystyle \frac{x_{\alpha}}{\sqrt{2}}} \overline{D}_{\alpha \beta}
\!=\!
\left( \!
{\displaystyle \frac{x^\dag}{\sqrt{2}}} F
 \!+\!
{\displaystyle \frac{x^{\mbox{\scriptsize T}}}{\sqrt{2}}} \overline{D} \!
\right)_{\!\!\beta} ,~~
\overline{[\alpha\gamma|\beta\delta]}
\!=\!
[\alpha\beta|\gamma\delta] ,
\EA
\label{relationFhatF}
\eeqa\\[-8pt]
in which all the terms except the $F$ and $\overline{D}$ terms
are canceled with each other.
The relations
$
\left( \!
\widehat{F} {\displaystyle \frac{x}{\sqrt{2}}}
\!-\!
\widehat{D} {\displaystyle \frac{\overline{x}}{\sqrt{2}}} \!
\right)_{\!\!\beta}
\!=\!
\left( \!
F {\displaystyle \frac{x}{\sqrt{2}}}
\!-\!
D {\displaystyle \frac{\overline{x}}{\sqrt{2}}} \!
\right)_{\!\!\beta}
$
etc. are also proved similarly.

In the matrix representation for $\slashed{\cal F}$,
further we have a simplified expression such as
\beqa
g_x \!\!
\left[ \!\!\!
\BA{cc} 
-M x^{\mbox{\scriptsize T}} & M x^\dag \\
\\[-2pt]
-\overline{M} x^{\mbox{\scriptsize T}} & \overline{M} x^\dag 
\EA \!\!\!
\right]
\!\!+\!\!
\left[ \!\!\!
\BA{cc} 
- \overline{x} M^{\dag} & - \overline{x} M^{\mbox{\scriptsize T}} \\
\\[-2pt]
x M^{\dag} & x M^{\mbox{\scriptsize T}} 
\EA \!\!\!
\right] \!\!
g^\dag_x
\!=\!
- \!
\left[ \!\!\!
\BA{cc} 
\overline{x} M^{\dag} \!+\! M x^{\mbox{\scriptsize T}} &
\overline{x} M^{\mbox{\scriptsize T}} \!-\! M x^\dag \\
\\[-2pt]
- \overline{( \overline{x} M^{\mbox{\scriptsize T}} \!-\! M x^\dag )} &
- \overline{( \overline{x} M^{\dag} \!+\! M x^{\mbox{\scriptsize T}})}
\EA \!\!\!
\right] 
\!\equiv\!
\widehat{\slashed{\cal F}}_{\!\!M}.
\label{gxMxgx}
\eeqa\\[-10pt]
At last,
we can reach the final form for
$\slashed{\cal F}$
more simply expressed as\\[-16pt]  
\beqa
\!\!\!\!\!\!\!\!\!\!
\BA{ll}
&\slashed{\cal F}
\!\!=\!\!
\left[ \!\!\!\!
\BA{ccc}
&\!\!\!\!\!\!\!\!\!\!\!\!\!\!
\widehat{\slashed{\cal F}} 
\!+\!
\widehat{\slashed{\cal F}}_{\!\!M}
&
\!\!\!\!\!\!\!\!
\left[ \! 
g_x \widehat{{\cal F}}
\!\!+\!\!
M^{\! \mbox{\scriptsize T}} \! x
\!\!-\!\!
M^{\! \dag} \overline{x} \!
\right] \!\!
\left[ \!\!\!
\BA{c}
{\displaystyle \frac{\overline{x}}{\sqrt{\! 2}}} \\
\\[-12pt]
{\displaystyle-\frac{x}{\sqrt{\! 2}}}
 \EA \!\!\!
\right] \!
\!+\!
z g_x \!\!
\left[ \!\!\!
\BA{c}
\sqrt{ 2} M \\
\\[4pt]
\sqrt{ 2} \overline{M}
\EA \!\!\!
\right] 
\\[6pt]
&
\!\!\!\!
\left[ \!
{\displaystyle \frac{x^{\! \mbox{\scriptsize T}}}{\sqrt{\! 2}}}
~~{\displaystyle-\frac{x^{\! \dag}}{\sqrt{\! 2}}} \!
\right] \!\!
\left[ \!
\widehat{{\cal F}} g^{ \dag}_x 
\!\!+\!\!
x^{\dag} \overline{M}
\!\!-\!\!
x^{\!\mbox{\scriptsize T}} \! M \!
\right] \!
\!\!+\!\!
z \!
\left[ \! \sqrt{ 2} \! M^{\dag} ~\! 
\sqrt{ 2} \! M^{\! \mbox{\scriptsize T}} \!
\right] \! g^{\dag}_x 
& 0
\EA \!\!\!
\right] \!\! .
\EA
\label{transF3}
\eeqa
The SCF parameters
$M$ originally has
any no correlations with motions of
the paired mode variables
$a$ and $b$.
Thus, from the physical viewpoint,
it is natural to make a theoretical set up
under the condition
produced by the following constraint:\\[-16pt]
\beqa
\BA{l}
g_x \widehat{{\cal F}} \!
\left[ \!\!
\BA{c} 
~{\displaystyle \frac{\overline{x}}{\sqrt{\! 2}}} \\
\\[-8pt]
{\displaystyle-\frac{x}{\sqrt{\! 2}}} 
\EA \!\!
\right] \!
+
z g_x \!
\left[ \!\!
\BA{c}
\sqrt{ 2} M \\
\\[12pt]
\sqrt{ 2} \overline{M}
\EA \!\!
\right] 
\!=\!
-
\left[ 
M^{ \mbox{\scriptsize T}} \! x
\!-\!
M^{\dag} \overline{x} 
\right] \!\!
\left[ \!\!
\BA{c} 
~{\displaystyle \frac{\overline{x}}{\sqrt{\! 2}}} \\
\\[-8pt]
{\displaystyle-\frac{x}{\sqrt{\! 2}}} 
\EA \!\!
\right] \! ,
\EA
\label{transF4}
\eeqa
multiplying which by
$
\left[ \!
{\displaystyle \frac{x^{\! \mbox{\scriptsize T}}}{\sqrt{\! 2}}}
~~~~~
{\displaystyle-\frac{x^{\! \dag}}{\sqrt{\! 2}}} \!
\right] \!
g^{-1}_x 
$
from the left,
we have\\[-8pt]
\beqa
\!\!\!\!\!\!
\BA{l}
\left[ \!
{\displaystyle \frac{x^{\! \mbox{\scriptsize T}}}{\sqrt{\! 2}}}
~~~
{\displaystyle-\frac{x^{\! \dag}}{\sqrt{\! 2}}} \!
\right] \!
\widehat{{\cal F}} \!
\left[ \!\!
\BA{c} 
~{\displaystyle \frac{\overline{x}}{\sqrt{\! 2}}} \\
\\[-8pt]
{\displaystyle-\frac{x}{\sqrt{\! 2}}} 
\EA \!\!
\right] \!
\!+\!
z \!
\left[ \!
{\displaystyle \frac{x^{\! \mbox{\scriptsize T}}}{\sqrt{\! 2}}}
~~~
{\displaystyle-\frac{x^{\! \dag}}{\sqrt{\! 2}}} \!
\right] \!\!
\left[ \!\!
\BA{c}
\sqrt{ 2} M \\
\\[12pt]
\sqrt{ 2} \overline{M}
\EA \!\!
\right] \!
\!=\!
-\!
\left[ 
M^{ \mbox{\scriptsize T}} \! x
\!-\!
M^{\dag} \overline{x} 
\right] \!\!
\left[ \!
{\displaystyle \frac{x^{\! \mbox{\scriptsize T}}}{\sqrt{\! 2}}}
~~~
{\displaystyle-\frac{x^{\! \dag}}{\sqrt{\! 2}}} \!
\right] \! 
g^{\!-1}_x \!\!
\left[ \!\!
\BA{c} 
~{\displaystyle \frac{\overline{x}}{\sqrt{\! 2}}} \\
\\[-8pt]
{\displaystyle-\frac{x}{\sqrt{\! 2}}} 
\EA \!\!
\right] \! .
\EA
\label{transF5}
\eeqa\\[-10pt]
In the above,
using the result
(\ref{xdagFx}),
we obtain a relation\\[-20pt]
\beqa
\BA{l}
(1\!+\! z) \!
\left[ 
M^{ \mbox{\scriptsize T}} \! x
\!-\!
M^{\dag} \overline{x} 
\right] \!\!
\left[ \!
{\displaystyle \frac{x^{\! \mbox{\scriptsize T}}}{\sqrt{\! 2}}}
~~~~~
{\displaystyle-\frac{x^{\! \dag}}{\sqrt{\! 2}}} \!
\right] \! 
g^{-1}_x \!\!
\left[ \!
\BA{c} 
~{\displaystyle \frac{\overline{x}}{\sqrt{\! 2}}} \\
\\[-8pt]
{\displaystyle-\frac{x}{\sqrt{\! 2}}} 
\EA \!
\right] \!
=
0 .
\EA
\label{transF5}
\eeqa\\[-8pt]
Thus we get a relation
$
M^{\! \mbox{\scriptsize T}} \! x
\!\!-\!\!
M^{\! \dag} \overline{x}
\!\!=\!\!
0
$ 
which strongly affects on a role of
the SCF parameters M.

\newpage

On the other hand, 
following to the form of the GDM
(\ref{anotherGDM}),
we introduce a new quantity,\\[2pt]
$
{\displaystyle \frac{1}{2}}\widehat{\slashed{W}}
\!\equiv\!
{\displaystyle \frac{1}{2}} 
\left[ \!
\BA{cc} 
\rho - {\displaystyle \frac{1}{2}} \!\cdot\! 1_{N} & - \overline{\kappa} \\
\\[-16pt]
\kappa &\!\!\!\! - \overline{\rho} +
{\displaystyle\frac{1}{2}} \!\cdot\! 1_{N} \!
\EA \!
\right]
$
which is transformed as\\[-2pt]
\beqa
\BA{l}
{\displaystyle \frac{1}{2}}
\widehat{\slashed{W}}
=
{\displaystyle\frac{1}{2}}
\slashed{W}
-
{\displaystyle\frac{1}{2}}
{\displaystyle\frac{1}{1 \!+\! z}} \!
\left[ 
{\displaystyle\frac{1}{2}}
\slashed{W} \!
\left[ \!
\BA{cc} 
{\displaystyle \frac{\overline{x}}{\sqrt{\! 2}}}
{\displaystyle \frac{x^{\! \mbox{\scriptsize T}}}{\sqrt{\! 2}}} &
- 
{\displaystyle \frac{\overline{x}}{\sqrt{\! 2}}}
{\displaystyle \frac{x^{\! \dag}}{\sqrt{\! 2}}} \\
\\[-6pt]
-
{\displaystyle \frac{x}{\sqrt{\! 2}}}
{\displaystyle \frac{x^{\! \mbox{\scriptsize T}}}{\sqrt{\! 2}}} &
{\displaystyle \frac{x}{\sqrt{\! 2}}}{\displaystyle\frac{x^{\! \dag}}{\sqrt{\! 2}}}
\EA \!
\right] 
\!+\!
\left[ \!
\BA{cc} 
{\displaystyle \frac{\overline{x}}{\sqrt{\! 2}}}
{\displaystyle \frac{x^{\! \mbox{\scriptsize T}}}{\sqrt{\! 2}}} &
- 
{\displaystyle \frac{\overline{x}}{\sqrt{\! 2}}}
{\displaystyle \frac{x^{\! \dag}}{\sqrt{\! 2}}} \\
\\[-6pt]
-
{\displaystyle \frac{x}{\sqrt{\! 2}}}
{\displaystyle \frac{x^{\! \mbox{\scriptsize T}}}{\sqrt{\! 2}}} &
{\displaystyle \frac{x}{\sqrt{\! 2}}}{\displaystyle\frac{x^{\! \dag}}{\sqrt{\! 2}}} \!
\EA \!
\right] 
{\displaystyle\frac{1}{2}}
\slashed{W} 
\right] .
\EA
\label{anotherGDM2}
\eeqa\\[2pt]
With the aid of the relation
$
g^2_x
\!=\!\!
\left[ \!\!\!
\BA{cc} 
1_{N} -
{\displaystyle \frac{\overline{x}}{\sqrt{2}}}
{\displaystyle \frac{x^{\mbox{\scriptsize T}}}{\sqrt{2}}}&
{\displaystyle \frac{\overline{x}}{\sqrt{2}}}
{\displaystyle \frac{x^\dag}{\sqrt{2}}} \\
\\[-14pt]
{\displaystyle \frac{x}{\sqrt{2}}}
{\displaystyle \frac{x^{\mbox{\scriptsize T}}}{\sqrt{2}}}&
1_{N} - 
{\displaystyle \frac{x}{\sqrt{2}}}
{\displaystyle \frac{x^\dag}{\sqrt{2}}}
\EA \!\!\!
\right] \!\! ,
$
the ${\displaystyle \frac{1}{2}}\widehat{\slashed{W}}$
is simply rewritten as\\[-2pt]
\beqa
\BA{l} 
{\displaystyle \frac{1}{2}}
\widehat{\slashed{W}}
=
{\displaystyle \frac{1}{2}} 
\left[ \!
\BA{cc} 
\rho - {\displaystyle \frac{1}{2}} \!\cdot\! 1_{N} & - \overline{\kappa} \\
\\[-16pt]
\kappa &\!\! - \overline{\rho} +
{\displaystyle\frac{1}{2}} \!\cdot\! 1_{N} \!
\EA 
\right] 
\!=
{\displaystyle\frac{1}{2}}
\slashed{W}
-
{\displaystyle\frac{1}{2}}
{\displaystyle\frac{1}{1 \!+\! z}} 
\left[ 
\slashed{W}
\!-\!
{\displaystyle\frac{1}{2}} 
\left( 
\slashed{W} \! g^2_x 
\!+\!
g^2_x 
\slashed{W} 
\right)
\right] .
\EA
\label{anotherGDM3}
\eeqa\\[-8pt]
Using again the matrix ${\cal F}$
(\ref{calF}),
parallel to the above formula for 
${\displaystyle \frac{1}{2}} \widehat{\slashed{W}}$,
the matrix $\widehat{\slashed{\cal F}}$ defined
in the last line of
(\ref{transbygx})
is also simply changed to\\[-10pt]
\beqa
\BA{l} 
\widehat{\slashed{\cal F}}
=
\left[ 
\BA{cc} 
\widehat{\slashed{F}} & \widehat{\slashed{D}} \\
\\[-8pt]
- \overline{\widehat{\slashed{D}}} & - \overline{\widehat{\slashed{F}}} 
\EA 
\right] 
=
\widehat{{\cal F}}
-
{\displaystyle\frac{1}{1 \!+\! z}} 
\left[ 
2 {\cal F}
-
\left( 
{\cal F} \! g^2_x 
+
g^2_x 
{\cal F}
\right)
\right] ,~
\widehat{\cal F}
\equiv
\left[ 
\BA{cc} 
\widehat{F}&  \widehat{D} \\
\\[-6pt]
- \overline{\widehat{D}} & - \overline{\widehat{F}} 
\EA 
\right] .
\EA
\label{hatcalF}
\eeqa\\[-4pt]
After complicated calculations, 
$\widehat{\cal F}$
is expressed by ${\cal F}$, 
two block matrices in component of GDM 
$\slashed{W}$
given by
(\ref{anotherGDM}), i.e.,
$R \!-\! {\displaystyle \frac{1}{2}} \!\cdot\! 1_{N}$ and $K$,
and unpaired mode amplitudes $x$ and $\overline{x}$
in a long form as\\[-8pt]
\beqa
\!\!\!\!\!\!\!\!\!\!
\BA{ll}
& 
\widehat{\cal F}
\!=\!
{\cal F} \\
\\[-10pt]
&\!\!\!\!\!\!\!\!
-
{\displaystyle\frac{1}{1 \!+\! z}} \!\!
\left[ \!\!
\BA{cc}
~~~~
[\alpha \beta | \gamma \delta] \!
\left[ \!
R \!-\! {\displaystyle \frac{1}{2}} \!\cdot\! 1_{\!N}
~~~~\!
- \overline{K} \!
\right] \!\!
\left[ \!\!\!
\BA{c}
~{\displaystyle \frac{\overline{x}}{\sqrt{\! 2}}}
{\displaystyle \frac{x^{\! \mbox{\scriptsize T}}}{\sqrt{\! 2}}}  \\
\\[-6pt]
-
{\displaystyle \frac{x}{\sqrt{\! 2}}}
{\displaystyle \frac{x^{\! \mbox{\scriptsize T}}}{\sqrt{\! 2}}}
\EA \!\!\!
\right] 
&
\!-
{\displaystyle \frac{1}{2}} 
[\alpha \gamma | \beta \delta] \!
\left[ \!
K
~~~~~ 
\!-\!
\overline{R} \!+\! {\displaystyle \frac{1}{2}} \!\cdot\! 1_{\!N} \!
\right] \!\!
\left[ \!\!\!
\BA{c}
~{\displaystyle \frac{\overline{x}}{\sqrt{\! 2}}}
{\displaystyle \frac{x^{\! \mbox{\scriptsize T}}}{\sqrt{\! 2}}}  \\
\\[-6pt]
-
{\displaystyle \frac{x}{\sqrt{\! 2}}}
{\displaystyle \frac{x^{\! \mbox{\scriptsize T}}}{\sqrt{\! 2}}}
\EA \!\!\!
\right]  \\
\\[-6pt]
\!-
{\displaystyle \frac{1}{2}} 
\overline{[\alpha \gamma | \beta \delta]} \!
\left[ \!
R \!-\! {\displaystyle \frac{1}{2}} \!\cdot\! 1_{\!N}
~~~~\!
- \overline{K} \!
\right] \!\!
\left[ \!\!\!
\BA{c}
-{\displaystyle \frac{\overline{x}}{\sqrt{\! 2}}}
{\displaystyle \frac{x^{\! \dag}}{\sqrt{\! 2}}}  \\
\\[-6pt]
~
{\displaystyle \frac{x}{\sqrt{\! 2}}}
{\displaystyle \frac{x^{\! \dag}}{\sqrt{\! 2}}}
\EA \!\!\!
\right]
&
~~~~\!
\overline{[\alpha \beta | \gamma \delta]} \!
\left[ \!
K
~~~~~ 
\!-\!
\overline{R} \!+\! {\displaystyle \frac{1}{2}} \!\cdot\! 1_{\!N} \!
\right] \!\!
\left[ \!\!\!
\BA{c}
-{\displaystyle \frac{\overline{x}}{\sqrt{\! 2}}}
{\displaystyle \frac{x^{\! \dag}}{\sqrt{\! 2}}}  \\
\\[-6pt]
~
{\displaystyle \frac{x}{\sqrt{\! 2}}}
{\displaystyle \frac{x^{\! \dag}}{\sqrt{\! 2}}}
\EA \!\!\!
\right]
\EA \!\!\!
\right] \\
\\
&\!\!\!\!\!\!\!\!
-
{\displaystyle\frac{1}{1 \!+\! z}} \!\!
\left[ \!\!
\BA{cc}
~~~~
[\alpha \beta | \gamma \delta] \!\!
\left[ \!
{\displaystyle \frac{\overline{x}}{\sqrt{\! 2}}}
{\displaystyle \frac{x^{\! \mbox{\scriptsize T}}}{\sqrt{\! 2}}}
~~
 -\! {\displaystyle \frac{\overline{x}}{\sqrt{\! 2}}}
{\displaystyle \frac{x^{\! \dag}}{\sqrt{\! 2}}} \!
\right] \!\!\!
\left[ \!\!\!
\BA{c}
R \!-\! {\displaystyle \frac{1}{2}} \!\cdot\! 1_{\!N}  \\
\\[-6pt]
K
\EA \!\!\!\!
\right] 
&
\!-
{\displaystyle \frac{1}{2}} 
[\alpha \gamma | \beta \delta] \!\!
\left[ \!
 - {\displaystyle \frac{x}{\sqrt{\! 2}}}
{\displaystyle \frac{x^{\! \mbox{\scriptsize T}}}{\sqrt{\! 2}}}
~~~\!\!
{\displaystyle \frac{x}{\sqrt{\! 2}}}
{\displaystyle \frac{x^{\! \dag}}{\sqrt{\! 2}}} \!
\right] \!\!\!
\left[ \!\!\!
\BA{c}
R \!-\! {\displaystyle \frac{1}{2}} \!\cdot\! 1_{\!N}  \\
\\[-6pt]
K
\EA \!\!\!\!
\right]  \\
\\[-6pt]
\!-
{\displaystyle \frac{1}{2}} 
\overline{[\alpha \gamma | \beta \delta]} \!\!
\left[ \!
{\displaystyle \frac{\overline{x}}{\sqrt{\! 2}}}
{\displaystyle \frac{x^{\! \mbox{\scriptsize T}}}{\sqrt{\! 2}}}
~~\!\!\!
- 
{\displaystyle \frac{\overline{x}}{\sqrt{\! 2}}}
{\displaystyle \frac{x^{\! \dag}}{\sqrt{\! 2}}} \!
\right] \!\!\!
\left[ \!\!\!\!
\BA{c}
- \overline{K} 
\\
\\[-6pt]
-\overline{R} \!+\! {\displaystyle \frac{1}{2}} \!\cdot\! 1_{\!N}
\EA \!\!\!\!
\right]
&
~~~\!
\overline{[\alpha \beta | \gamma \delta]} \!\!
\left[ \!
- {\displaystyle \frac{x}{\sqrt{\! 2}}}
{\displaystyle \frac{x^{\! \mbox{\scriptsize T}}}{\sqrt{\! 2}}}
~~~
{\displaystyle \frac{x}{\sqrt{\! 2}}}
{\displaystyle \frac{x^{\! \dag}}{\sqrt{\! 2}}} \!
\right] \!\!\!
\left[ \!\!\!\!
\BA{c}
- \overline{K}  \\
\\[-6pt] 
-\overline{R} \!+\! {\displaystyle \frac{1}{2}} \!\cdot\! 1_{\!N} 
\EA \!\!\!\!
\right]
\EA \!\!\!\!
\right] \! .
\EA
\label{hatcalF2}
\eeqa
As for the relation $g^2_x$,
we notice that the relation is
rewritten as\\[-12pt]
\beqa
\BA{l} 
g^2_x
-
1_{2N}
=
\left[ \!\!
\BA{cc} 
 -
{\displaystyle \frac{\overline{x}}{\sqrt{2}}}
{\displaystyle \frac{x^{\mbox{\scriptsize T}}}{\sqrt{2}}}&
{\displaystyle \frac{\overline{x}}{\sqrt{2}}}
{\displaystyle \frac{x^\dag}{\sqrt{2}}} \\
\\[-6pt]
{\displaystyle \frac{x}{\sqrt{2}}}
{\displaystyle \frac{x^{\mbox{\scriptsize T}}}{\sqrt{2}}}&
 - 
{\displaystyle \frac{x}{\sqrt{2}}}
{\displaystyle \frac{x^\dag}{\sqrt{2}}}
\EA \!\!
\right] 
=
\left[ \!\!
\BA{c}
~~{\displaystyle \frac{\overline{x}}{\sqrt{ 2}}} \\
\\[-2pt]
-
{\displaystyle \frac{x}{\sqrt{ 2}}}
\EA \!\!
\right] \!
\left[ \!
- 
{\displaystyle \frac{x^{ \mbox{\scriptsize T}}}{\sqrt{ 2}}}
~~~~
{\displaystyle \frac{x^{ \dag}}{\sqrt{ 2}}} \!
\right] ,
\EA
\label{squaregx-1Matrix}
\eeqa\\[-6pt] 
through which
we get a relation 
$
\left(  g^2_x - 1_{2N}  \right)^2
\!=\!
-  \left( 1 - z^2  \right)
\left(  g^2_x - 1_{2N}  \right)
$.
Then we have an equation for OP, namely,
$
\left(  g^2_x - z^2 \!\cdot\! 1_{2N} \right)
\left( g^2_x - 1_{2N} \right)
=
0
$.
From now on we will restrict the Hilbert space to
a subspace $| >$ in which the OP, 
$g^2_x - z^2 \!\cdot\! 1_{2N}$, satisfies
the relation,
$\left(  g^2_x - z^2 \!\cdot\! 1_{2N} \right) | > ~\!\!\! = 0$.

Under the equation
(\ref{transF3})
and the constraint
(\ref{transF4})
and the relation
(\ref{hatcalF}),
thus we reach our final goal of deriving the following
modified
$SO(2N \!+\! 1)$ HB EE with
$\widehat{{\cal F}}$
(\ref{hatcalF2}):
\beqa
\left.
\BA{ll}
& 
\left\{ 
\widehat{{\cal F}} 
\!-\!
2 (1 \!-\! z)
{\cal F}
- \!
\left[ \!
\BA{cc} 
\overline{x} M^{\dag} \!+\! M x^{\mbox{\scriptsize T}} &
\overline{x} M^{\mbox{\scriptsize T}} \!-\! M x^\dag \\
\\[-2pt]
- \overline{( \overline{x} M^{\mbox{\scriptsize T}} \!-\! M x^\dag )} &
- \overline{( \overline{x} M^{\dag} \!+\! M x^{\mbox{\scriptsize T}})}
\EA \!
\right] \!
\right\} \!
\left[ \!
\BA{c}
a \\
\\[-2pt]
b
\EA \!
\right]_i
\!=\!
\varepsilon_i \!
\left[ \!
\BA{c}
a \\
\\[-2pt]
b
\EA \!
\right]_i \! , \\
\\[-2pt]
&
\widehat{{\cal F}} \!
\left[ \!\!
\BA{c} 
~{\displaystyle \frac{\overline{x}}{\sqrt{\! 2}}} \\
\\[-10pt]
{\displaystyle-\frac{x}{\sqrt{\! 2}}} 
\EA \!\!
\right] \!
+
z \!
\left[ \!\!
\BA{c}
\sqrt{ 2} M \\
\\[14pt]
\sqrt{ 2} ~\! \overline{M}
\EA \!\!
\right]
=
0 ,~~
M^{ \mbox{\scriptsize T}} x
-
M^{ \dag} ~\! \overline{x}
=
0 ,
\EA 
\right\}
\label{transbygx3}
\eeqa\\[-4pt]
where $i$ means a {\bf QP} state
and the $SO(2N \!\!+\!\! 1)$ GHB
MF OP
$\widehat{\cal F}$
(\ref{hatcalF})
has with a complicated form of interaction term. 
The above type of equations with slight modifications were already
obtained in 
\cite{NishiProvi.12}
in which, 
of course, 
the original HB OP
is replaced by
${\cal F}$
instead of
$\widehat{\cal F}$.
Both the equations in the second line of
(\ref{transbygx3})
are similar to those given by
(\ref{eigeneq3})
in which, however, it must be paid attention to
a role of the unpaired mode amplitudes.
The $x$ and $\overline{x}$
are exchanged each other
in both the equations in the second line of
(\ref{transbygx3}).
In deriving
the modified
$SO(2N \!+\! 1)$ HB EE,
the column vector $\sqrt{2}\sqrt{1 \!+\! z}  L$ given by
the relation
(\ref{gxWgx2})
plays an important role in each step of calculative process.
In closing this section,
we finally attract a significant attention 
concerning a basic problem what is the physical meaning of the quantity,
$\! - \! \sqrt{2}\! \sqrt{1 \!+\! z} L$.$\!\!$
\def\erw#1{{<\!#1\!>_G}}
Noticing the WF
$\ket G \!=\! U(G) \ket 0$, i.e.,
$SO(2N \!+\! 1)$ CS rep, 
and using the relations
$
A_{\alpha \beta }
\!=\!
a_{\alpha \beta }
\!-\!
{\displaystyle \frac{\overline{x}_\alpha y_\beta }{2(1 \!+\! z)}} 
$,
$
B_{\alpha \beta }
\!=\!
b_{\alpha \beta }
\!+\!
{\displaystyle \frac{x_\alpha y_\beta }{2(1 \!+\! z)}}
$
and
$
y_\beta
\!=\!
a_{\gamma \beta } x_\gamma
\!-\!
b_{\gamma \beta } \overline{x}_\gamma
$,
the {\bf QP} vacuum EV
of the annihilation OP $c_\alpha$, i.e.,
$\erw{c_\alpha }$
is found mathematically into the following form:\\[-16pt]
\beqa
\BA{ll}
\erw{c_\alpha }
&\!\!\!
=
{\displaystyle \frac{1}{2}} \!
\left(
\overline{A}_{\alpha \beta} y_{\beta}
\!+\!
B_{\alpha \beta} \overline{y}_{\beta}
\right)
=
{\displaystyle \frac{1}{2}} \!
\left\{ \!
\left( \!
a_{\alpha \beta }
\!-\!
{\displaystyle \frac{\overline{x}_\alpha y_\beta }{2(1 \!+\! z)}} \!
\right) \!
y_{\beta}
\!+\!
\left( \!
b_{\alpha \beta }
\!+\!
{\displaystyle \frac{x_\alpha y_\beta }{2(1 \!+\! z)}} \!
\right) \!
\overline{y}_{\beta} \!
\right\} \\
\\[-8pt]
&\!\!\!
=
{\displaystyle \frac{1}{2}} \!
\left(
\overline{a}_{\alpha \beta} y_{\beta}
\!+\!
b_{\alpha \beta} \overline{y}_{\beta}
\right)
=
\left( \!
{\displaystyle \frac{1}{2}}
\delta_{\alpha \beta } \!-\! \overline{R}_{\alpha \beta } \!
\right) \!
x_\beta
\!-\!
K_{\alpha \beta }\overline{x}_{\beta } 
=
-\sqrt{2}\sqrt{1 \!+\! z} L_\alpha ,
\EA
\eeqa\\[-10pt]
due to
(\ref{gxWgx2})
and
$
\erw{c^\dagger_\alpha}
=
\erw{\overline{c}_\alpha }
$.
Thus,
the physical meaning of the quantity,
$-\sqrt{2}\sqrt{1 \!+\! z} L_\alpha$,
is made clear. 
Thus, it is said that the quantity is the {\bf QP} vacuum EV
of annihilation OP $c_\alpha$.

Originally,
the $SO(2N \!+\! 1)$ Hamiltonian
(\ref{mean-field Hamiltonian}), i.e.,
$\!H_{SO(2N \!+\! 1)}$
with $\!M \!\!\neq\! 0$
can not be essentially regarded as an SCF Hamiltonian. 
Such a situation is in very contrast to the fact that
theoretical self-consistent MF Hamiltonian 
is obtained by through the appropriate averaging of
the original HB ($SO(2N)$) Hamiltonian 
over the fermion states. 
As a matter of fact,
at the early stage of the present work, 
it has already been pointed out that
the unknown parameters $k$ and $l$ involved in $M$ 
are considered to connect strongly with the constraint
to select out the physical fermion space 
from the whole Hilbert space
\cite{FYN.77}.
This gives a self-consistency problem for 
the unknown parameter.
In the succeeding section,
we will discuss such a stubborn and trouble problem.


\newpage

\setcounter{equation}{0}
\renewcommand{\theequation}{\arabic{section}.\arabic{equation}}

\section{Discussions, perspective and summary}

~~
The embedding of the $SO(2N \!+\! 1)$ group into 
an $SO(2N \!+\! 2)$ group
leads us to an unified formulation by which
we can treat 
the paired and unpaired mode amplitudes with an equal footing.
Let us define 
$(N \!+\! 1) \!\times\! (N \!+\! 1)$ matrices 
${\cal A}$ and ${\cal B}$
accompanying with
$y \!=\! x^{\mbox{\scriptsize T}} \! a \!-\! x^{\dag }b$
and introduce a new $SO(2N \!+\! 2)$ matrix
${\cal G}$
similar to the $SO(2N)$ matrix
(\ref{RepBogotrans}) 
as follows:\\[-8pt]
\beq
{\cal A}
\!\equiv\!
\left[ \!\!
\BA{cc}
A & {\displaystyle -\frac{\overline{x}}{2}} \\
\\[-12pt]
{\displaystyle \frac{y}{2}} &{\displaystyle \frac{1\!+\!z}{2}}
\EA \!\!
\right],
~~
{\cal B}
\!\equiv\!
\left[ \!\!
\BA{cc}
B & {\displaystyle \frac{x}{2}} \\
\\[-12pt]
{\displaystyle -\frac{y}{2}} &{\displaystyle \frac{1\!-\!z}{2}}
\EA \!\!
\right] ,~~
{\cal G}
\!\equiv\!
\left[ \!
\BA{cc}
{\cal A} & \overline{\cal B} \\
\\[4pt]
{\cal B} & \overline{\cal A}
\EA \!
\right],
\label{calAcalBcalG}
\eeq\\[-6pt]
As the ortho-normalization for the matrix $G$ holds,
i.e.,
$
G^{\dag } G
\!\!=\!\!
G G^{\dag }
\!\!=\!\! 1_{2N \!+\! 1} ~
\mbox{and}~\!
\det G \!\!=\!\! 1,
(\ref{UGdetG})
$,
${\cal A}$ and ${\cal B}$
also satisfy the ortho-normalization condition
$
{\cal G}^{\dag } \! {\cal G}
\!=\!
{\cal G} \! {\cal G}^{\dag }
\!=\! 1_{2N \!+\! 2}~
\mbox{and}
\det {\cal G} \!=\! 1 \!
$ 
\cite{Fuk.77,Fuk.81}.\\
Parallel to the previous work$\!$
\cite{Fuk.77},
variables ${\cal A}, {\cal B}$ and $y$ are bosonized as
$
\hbox{\boldmath ${\cal A}$},
\hbox{\boldmath ${\cal B}$}
$
and
$\!\hbox{\boldmath ${\cal Y}$}\!$,
respectively,
according to the same manner as the one
adopted in Appendix A of the recent our work
\cite{SeiyaJoao.15}.

In the boson images of 
$SO(2N \!+\! 1)$ Lie OPs,
particularly
the images $\hbox{\mathbf c}_\alpha$ and
$\hbox{\mathbf c}_\alpha^\dagger$
must satisfy the anti-commutation relations,
$
\{\hbox{\mathbf c} _\alpha, \hbox{\mathbf c}^\dagger_\beta\}
\!=\! \delta_
{\alpha\beta },~
\{\hbox{\mathbf c}_\alpha, \hbox{\mathbf c}_\beta\}
\!=\! 0~
\mbox{and}~\!
\{\hbox{\mathbf c}^\dagger _\alpha, \hbox{\mathbf c}^\dagger _\beta\}
\!=\! 0
$ 
when they are
operated onto the spinor subspace within the whole space.
Therefore, here
we should adopt, 
in the underlying regular representation space,
the following boson image of the Hamiltonian
in which the Lagrange multiplier terms are added
to select out the controversial spinor subspace:\\[-18pt]
\beqa
\!\!\!\!
\left.
\BA{lll}
\hbox{\mathbf H}
\!\!\!&\!\!=\!\!&\!\!\!
h_{\alpha \beta } \!
\left( \!
\hbox{\mathbf E}^{\,\alpha }_{~\beta }
\!+\!
{\displaystyle \frac{1}{2}} \delta_{\alpha \beta } \!
\right)
\!+\!
{\displaystyle \frac{1}{4}} [\alpha\beta|\gamma\delta]
\left( \!
\{
\hbox{\mathbf E}^{\,\alpha }_{~\beta }
\!+\!
{\displaystyle \frac{1}{2}} \delta_{\alpha \beta },
\hbox{\mathbf E}^{\,\gamma }_{~\delta }
\!+\!
{\displaystyle \frac{1}{2}} \delta_{\gamma \delta }
\}
\!+\!
\{
\hbox{\mathbf E}^{\,\alpha \gamma },\hbox{\mathbf E}_{\delta \beta }
\} \!
\right)
\!+\!
\hbox{\mathbf H}^{~\!\prime } ,\\
\\[-12pt]
\hbox{\mathbf H}^{~\!\prime }
\!\!&\!\!=\!&\!\!
{\displaystyle \frac{1}{2}} 
\overline{k}_{\alpha \beta } 
\left( 
\{\hbox{\mathbf c}_\alpha,\hbox{\mathbf c}^\dagger_\beta\}
\!-\!
\delta_{\alpha \beta } 
\right)
\!+\!
{\displaystyle \frac{1}{4}}
l_{\alpha\beta }\{\hbox{\mathbf c}^\dagger_\alpha,
\hbox{\mathbf c}^\dagger_\beta\}
\!+\!
{\displaystyle \frac{1}{4}}
\overline{l}_{\alpha \beta }\{\hbox{\mathbf c}_\alpha,\hbox{\mathbf c}_
\beta\},~
(
\overline{k}_{\alpha \beta }
=
k_{\beta \alpha },~
l_{\alpha \beta }
=
l_{\beta \alpha }
) ,\\
\\[-10pt]
\hbox{\mathbf c}_\alpha
\!\!\!&\!=\!&\!\!\!
\sqrt{2} 
\left(
\hbox{\boldmath ${\cal A}$}^{\alpha\dagger }_r
\hbox{\boldmath ${\cal Y}$}_r
\!+\!
\hbox{\boldmath ${\cal Y}$}^\dagger_{\widetilde{r}}
\hbox{\boldmath ${\cal B}$}_{\alpha \widetilde{r}}
\right) , ( r \!=\! 1, \cdots, 2N\!+\!2),~
\hbox{\mathbf c}^\dagger_\alpha
\!=\!
- \overline{\hbox{\mathbf c}}_\alpha ,
\EA \!\!
\right\}
\label{Hamiltonianimage}
\eeqa\\[-14pt]
where
it happens that
the classical part of the Lagrange multiplier terms
corresponding to the origin of the SCF parameter $\!M\!$
arises naturally,
which is given as
$
M_{\!\alpha}
\!=\!\!
k_{\alpha \beta } \erw{c_\beta }
\!+\!
l_{\alpha \beta } \erw{c^\dagger_\beta }
$.
A possible determination of the unknown parameters
$k_{\alpha \beta}$ and $l_{\alpha \beta}$
has already been attempted, however, 
has been still incomplete yet
\cite{NishiProvi.12}.
This is a difficult problem to be solved completely.

Further
we point out that 
the present MFT is deeply related to the algebraic MFT
(AMFT)
which has been 
proposed by Rosensteel and Dancova 
\cite{Rosensteel.11,DancovaRosensteel.99,RosensteelDancova.98}
based on
the coadjoint orbit method
\cite{Kirillov.76}.
Following them,
when we try to apply the AMFT to the present theory,
every time
it is indispensably necessary to introduce an even-dimensional GDM 
on the $SO(2N \!+\! 2)$ CS rep
as is seen soon later.
So, we are demanded to prepare a 
$(2\!N \!+\! 2) \times (2\!N \!+\! 2)$ matrix
${\cal W}$ 
defined as \\[-8pt]
\beq
{\cal W}
\stackrel{\mathrm{def}}{=}
\left[ \!
\BA{cc} 
{\cal R} & {\cal K} \\
\\[-8pt]
-\overline{\cal K} & 1_{N + 1} - \overline{\cal R}
\EA \!
\right] 
=
{\cal W}^\dag ~
({\cal W}^2 = {\cal W}) ,
\left\{ \!
\BA{c}
{\cal R}
=
{\cal B}{\cal B}^\dag ,\\
\\[-8pt]
{\cal K}
=
{\cal B}{\cal A}^\dag .
\EA
\right.
\label{densitymat}
\eeq\\[-6pt]
The ${\cal W}$ is a natural extension of the GDM on the $SO(2N)$ CS rep
to the GDM on $SO(2N \!+\! 2)$ CS rep.
We should emphasized that
the HB GDMs on
the $SO(2\!N \!\!+\! 2) (\!\ni\! {\cal G})$ CS rep
are elements of the dual space ${\cal G}^*\!$ of the Lie algebra
$SO(2N \!+\! 2)$.
According to the basic ideas given in
\cite{RosensteelRowe.81,Rosensteel.81}
and
\cite{RoweRymanRosensteel.80},
we prepare 
both the HB GDM ${\cal W}$ and its coadjoint orbit
$
O_{\cal W}
=
\{{\cal G}{\cal W} {\cal G}^{-1}|~\! {\cal G} \! \in \! SO(2N \!+\! 2)\}
$.
In order to complete the geometrical picture of the $O_{\cal W}$,
we must establish the concept of the symplectic structure.
This is made to construct
a non-degenerate symplectic form $\omega$ at a point 
${\cal G}{\cal W} {\cal G}^{-1}$
 in the orbit $O_{\cal W}$ as
 an antisymmetric form defined on pairs of
 tangent vectors at
${\cal G}{\cal W} {\cal G}^{-1}$,\\[-10pt]
\beq
\omega_{~\!{\cal G}{\cal W} {\cal G}^{-1}}(X,~\!Y)
=
- i
\mbox{tr}
\left( {\cal G}{\cal W} {\cal G}^{-1} [X,~\!Y] \right) ,
\label{symplectic-form}
\eeq\\[-16pt]
where $\!X,Y \!\in\! SO(2N \!+\! 2)\!$
are tangental vectors at
${\cal G}{\cal W} {\cal G}^{-1}$.
The idempotent GDM ${\cal W}$ forms an orbit surface 
in the space of all GDMs that is invariant
with respect to the coadjoint group action,
$
{\cal W}
\!\rightarrow\!
{\cal G}{\cal W} {\cal G}^{-1}
$
for
$\!{\cal G} \!\in\! SO(2N \!+\! 2)$.
For both determinantal orbits and generic orbits,
the orbit surface is even-dimensional.
Thus, the dimension of the GDM necessarily becomes even-dimensional.
This is why we introduce the GDM on the $SO(2N \!\!+\! 2)$ CS rep 
instead of the GDM on the $SO(2N \!\!+\! 1)$ CS rep.
The determinantal orbit is, of course, 
regarded as a symplectic manifold.
The $SO(\! 2\!N \!+ \! 2 \!)$ canonical transformation of the $O_{\cal W}$
preserves the symplectic structure,\\[-14pt]
\beq
\omega_{~\!\cal W}(X,~\!Y)
=
\omega_{~\!{\cal G}{\cal W} {\cal G}^{-1}}
(\mbox{ad}_{\cal G}(X),~\!\mbox{~\!ad}_{\cal G}(Y)) ,~~
\mbox{~\!ad}_{\cal G}(X)
\equiv
{\cal G} X {\cal G}^{-1},~
\mbox{~\!ad}_{\cal G}(Y)
\equiv
{\cal G} Y {\cal G}^{-1} ,
\label{symplectic-form_calW}
\eeq\\[-14pt]
where $X,Y \!\in\! SO(2N \!+\! 2)$
are tangent vectors at ${\cal W}$.
While,
according to Rosensteel and Rowe again,
we also define
the coadjoint action $\mbox{ad}^*_{~\!\cal G}$ on the GDM
on the  $SO(2N \!+\! 2)$ CS rep
as, \\[-10pt]
\beq
\mbox{ad}^*_{~\!\cal G}({\cal W})
=
{\cal G}{\cal W} {\cal G}^{-1} .
\label{ad*W}
\eeq\\[-18pt]
The orbit surface $O_{\cal W}$ has
one-to-one correspondence with the coset space of
$SO(2N \!\!+\! 2)$ modulo.
Then, the isotropy subgroup arises
at the ${\cal W}$, say,
$
{\cal H}_{\cal W}
\!=\!
\{ 
h \!\in\! SO(2N \!\!+\!\! 2) ~\!\! | ~\!\! h{\cal W}h^{-1} 
\!=\! 
{\cal W} 
\}
$
and
the coset space is identified with the $O_{\cal W}$: 
$
\frac{SO(2N + 2)}{{\cal H}_{\cal W}}
\!\rightarrow\!
O_{\cal W},~
{\cal G} {\cal H}_{\cal W}
\!\rightarrow\!
{\cal G}{\cal W} {\cal G}^{-1} 
$.
\def\erw#1{{<\!#1\!>_{\cal G}}}
In the case of generic orbit, however,
the map  
$
U({\cal G}) \Phi
\!\rightarrow\!
{\cal G}{\cal W} {\cal G}^{-1}
$
is a many-to one correspondence.
We, here, have used the properties
$
U({\cal G}) \ket 0
\!=\!
U(G) \ket 0 
$
and
$
U(G) \ket 0
\!=\!
\bra 0 U(G) \ket 0
e^{\frac{1}{2}{\cal Q}_{pq}E^{pq}} \ket 0
$,
where ${\cal Q}_{pq}$
is given by
$
{\cal Q}_{pq}
\!=\!
({\cal B}{\cal A}^{-1})_{pq}
$
and
$E^{pq}$
belongs to the fermion $SO(2N \!+\! 2)$
Lie OPs
\cite{Fuk.81}.
To advance a construction of the AMFT on the
$SO(2N \!+\! 2)$ manifold,
\def\erw#1{{<\!#1\!>_G}}
we consider a model Hamiltonian $\widehat{H}$ on the 
$SO(2N \!+\! 2)$ manifold and
an energy function on $O_{\cal W}$
by a vacuum EV of the Hamiltonian as,\\[-22pt]
\ba
\!\!\!\!
\BA{l}
\widehat{H}_{SO(2N \!+\! 2)}
\!=\!
h_{pq } \!
\left( \!\!
E^p_{~q } 
\!+\! 
{\displaystyle \frac{1}{2}}\delta_{pq} \!
\right)
\!+\!
{\displaystyle \frac{1}{4}}[pq|st] E^{ps} \! E_{tq} ,~
H_{\!\cal W}({\cal G}{\cal W} {\cal G}^{-1})
\!\equiv
<\!U({\cal G})\Phi | \widehat{H}_{SO(2N \!+\! 2)}U({\cal G})\Phi \!> \! .
\EA
\label{Henergy-function}
\ea\\[-18pt]
Due to the many-to one correspondence,
the energy function on the $O_{\cal W}$ cannot take 
the vacuum EV of the model Hamiltonian.
This makes the evaluation
of the energy function to be ambiguous.
To remove such ambiguity,
we have a possibility to choice for the energy function as\\[-22pt]
\ba
{\cal H} ({\cal G}{\cal W} {\cal G}^{-1})
=
\mbox{min}_{~\!h \in H_{\cal W}} \!
<\!U({\cal G}) U(h) \Phi \,| H_{SO(2N \!+\! 2)}\,U({\cal G}) U(h) \Phi \!>,~
h \in H_{\cal W} .
\label{Henergy-function2}
\ea\\[-22pt]
The ground-state GDM lies on one of the orbits $O_{\cal W}$
and then the minimization for
the energy function must be made on the orbit surface
\cite{RosensteelRowe.81,Rosensteel.81}.
Thus, we unavoidably encounter a serious problem of
the determination of the orbit surface and
the exhaustive search for the energy minimum
at the same time.
Such a difficult problem will be studied elsewhere
in a near future.

As mentioned first,
to give a theory for collective motion
is still a currently topical problem.
There exists an alternative way,
the so-called 
\textquotedblleft {\bf moment map} $\!$\textquotedblright
method
proposed by Guillemin and Sternberg
\cite{GuilleminSternberg.80}
and by Ref.
\cite{RosensteelDancova.98}.
This method may be also considered to be a powerful tool
for constructing the microscopic theory to describe
motions with strongly collective correlations.
 
$\!\!\!$Finally, 
the essentials are summarized as follows: we have given 
the GHB-MFH in terms of the fermion 
$SO(2N \!\!+\!\! 1)$ Lie OPs.
We have diagonalized the GHB-MFH and
throughout the diagonalization of which,
we first have obtained the unpaired mode amplitudes
which are given by the SCF parameters appeared in the HBT
together with the additional SCF parameter
in the GHB-MFH
and by the $SO(2N \!\!+\!\! 1)$ group-parameter $z$.
Consequently, 
it turns out that
the magnitudes of these amplitudes are governed by such parameters. 
Then, we has made clear a new aspect of such the results
which have never been in the traditional works.
We also have studied the K\"{a}hler symmetric space 
$\frac{SO(2N)}{U(N)}$
and 
constructed a Killing potential
in the coset space $\frac{SO(2N)}{U(N)}$.
It is our great surprise that
the Killing potential is equivalent with the GHB GDM
$\widehat{\slashed{\cal W}}$
corresponding to the many-fermion state $\Phi$.
Generally speaking,
the GHB GDM is an important and useful tool to study 
the many-fermion problems.
See details in textbooks
\cite{BR.86}
and
\cite{Belyaev.68}. 
We further have developed vigorously another approach to
the fermion MFH based on
the different form of the GDM which is
essentially equivalent with the Killing potential.
In the ordinary course of things,
as a result,
we have derived the $SO(2N \!+\! 1)$ GHB MF OP
(\ref{hatcalF}) 
and
the modified HB EE
(\ref{transbygx3}) 
accompanying by
non-vanishing SCF parameters $M_\alpha$'s.
Thus,
we could give successfully a new fermion MFT
for the paired and unpaired modes
on the K\"{a}hler coset space
$\frac{SO(2N + 2)}{U(N + 1)}$
standing on the basic sprit of 
the algebraic mean-field viewpoint.

\newpage

\vspace{-0.5cm}
$\!\!\!\!\!\!\!\!\!${\bf Acknowledgements}
S. N. 
expresses his sincere thanks to
Professor Constan\c{c}a Provid\^{e}ncia for kind and
warm hospitality extended to
him at the Centro de F\'\i sica,
Universidade de Coimbra, Portugal.$\!$
This work was supported by FCT(Portugal) under the project
CERN/FP/83505/2008.

\vspace{-0.7cm}


\end{document}